\documentclass[a4paper,12pt]{article}

\usepackage{amsmath}\usepackage{amssymb}
\usepackage{latexsym}\usepackage{cite}

\usepackage{amsmath}
\usepackage[dvips]{graphicx}
\usepackage{pstricks}
\usepackage{color}
\usepackage{bm}

\usepackage{graphicx}

\definecolor{Red}{rgb}{1.00, 0.00, 0.00}
\definecolor{Green}{rgb}{0.00, 1.00, 0.00}
\definecolor{Blue}{rgb}{0.00, 0.00, 1.00}
\definecolor{Cyan}{rgb}{0.00, 1.00, 1.00}
\definecolor{Mymagenta}{rgb}{0.3, 0.00, 1.00}%
\definecolor{Magenta}{rgb}{1.00, 0.00, 1.00}
\definecolor{DeepSkyBlue}{rgb}{0.00, 0.75, 1.00}
\definecolor{DarkGreen}{rgb}{0.00, 0.39, 0.00}
\definecolor{SpringGreen}{rgb}{0.00, 1.00, 0.50}
\definecolor{Mygreen}{rgb}{0.00, 0.72, 0.00}
\definecolor{DarkOrange}{rgb}{1.00, 0.55, 0.00}
\definecolor{OrangeRed}{rgb}{1.00, 0.27, 0.00}
\definecolor{DeepPink}{rgb}{1.00, 0.08, 0.57}
\definecolor{DarkViolet}{rgb}{0.58, 0.00, 0.82}
\definecolor{SaddleBrown}{rgb}{0.57, 0.27, 0.07}
\definecolor{Black}{rgb}{1.00, 1.00, 1.00}
\definecolor{Ablue}{rgb}{0.10, 0.1, 1.00}

\newcommand{\be}{\begin{equation}}
\newcommand{\ee}{\end{equation}}
\newcommand{\tr}{\,\textup{tr}}
\newcommand{\ka}{\kappa}

\def\beq{\begin{equation}}
\def\eeq{\end{equation}}
\def\beqr{\begin{eqnarray}}
\def\eeqr{\end{eqnarray}}

\def\al{\alpha}
\def\bt{\beta}
\def\Ga{\Gamma}

\def\de{\delta}
\def\De{\Delta}

\def\ka{\kappa}
\def\si{\sigma}

\def\lam{\lambda}

\def\om{\omega}
\def\ep{\epsilon}

\def\sq{\sqrt}

\def\l{\left (}
\def\r{\right )}

\def\fr{\frac}
\def\la{\label}
\def\hs{\hspace}
\def\vs{\vspace}

\def\ov{\overline}
\def\tl{\tilde}
\def\tm{\times}

\def\lrar{\leftrightarrow}

\def\beqs{\begin{equation*}}
\def\eeqs{\end{equation*}}

\begin{document}

\begin{flushright}
November 23, 2016 \\
\end{flushright}

\vs{1cm}

\begin{center}
{\Large\bf

Calculable Cosmological CP Violation and

Resonant Leptogenesis

 %
 %
 }

\end{center}

\vspace{0.5cm}
\begin{center}
{\large
~Avtandil Achelashvili\footnote{E-mail: avtandil.achelashvili.1@iliauni.edu.ge},~ and
~Zurab Tavartkiladze\footnote{E-mail: zurab.tavartkiladze@gmail.com}
}
\vspace{0.5cm}

{\em Center for Elementary Particle Physics, ITP, Ilia State University, 0162 Tbilisi, Georgia}
\end{center}

\vspace{0.6cm}

\begin{abstract}

Within the extension of MSSM by two right handed neutrinos, which masses are degenerate at tree level,
we address the issue of leptogenesis. Investigating the quantum corrections in details, we show that the
lepton asymmetry is induced at 1-loop level and decisive role is played by the tau lepton Yukawa coupling.
On a concrete and predictive neutrino model, which enables to predict the CP violating $\delta $ phase and relate it
to the cosmological CP asymmetry, we demonstrate that the needed amount of the baryon asymmetry is generated via the
resonant leptogenesis.

\end{abstract}

\hspace{0.4cm}{\it Keywords:}~CP violation; Resonant Leptogenesis; Neutrino mass and mixing; Renormalization.

\hspace{0.4cm}PACS numbers:~11.30.Er, 98.80.Cq, 14.60.Pq, 11.10.Gh.

\section{Introduction}
\la{intro}

Simplest extension of the standard model (SM), required for accommodation  of the
atmospheric and solar neutrino data \cite{Gonzalez-Garcia:2014bfa}, is inclusion of the SM singlet
right handed neutrinos (RHN). The latter, having the Majorana mass, can generate neutrino masses via see-saw mechanism.
It is remarkable, that this simple construction also offers an elegant way for generating the baryon asymmetry of the Universe
through thermal leptogenesis
 \cite{Fukugita:1986hr} (for reviews see: \cite{{Giudice:2003jh},{Buchmuller:2004nz},{Davidson:2008bu}}).
 In order to reduce the number of parameters entering in the CP asymmetry, the minimalistic  approach with texture zeros
 has been put forward in Ref. \cite{Frampton:2002qc}. This approach enables one to relate the CP violating phase $\de $ (appearing
 in the neutrino oscillations) with the cosmological CP asymmetry
 \cite{{Frampton:2002qc},{Ibarra:2003up},{Shafi:2006nt},{Branco:2006hz},{Babu:2007zm},{Babu:2008kp},{Meroni:2012ze},{Harigaya:2012bw},
 {Ge:2010js},{Achelashvili:2016nkr}}.
Especially attractive looks the setup with two (or more) quasi-degenerate RHN's \cite{{Branco:2006hz},{Babu:2007zm},{Babu:2008kp},{Meroni:2012ze},{Achelashvili:2016nkr}}
 because, besides the further reduction of  the model parameter number, it offers the possibility for resonant leptogenesis
\cite{{Flanz:1996fb},{Pilaftsis:1997jf},{Pilaftsis:2003gt}} (for recent discussions on resonant leptogenesis see
\cite{{Blanchet:2012bk},{Dev:2015wpa},{Dev:2014laa},{Pilaftsis:2015bja}}).

With two degenerate RHNs, in \cite{Babu:2008kp}  all possible one texture zero $3\!\tm \!2$ Dirac type Yukawa couplings have been
investigated. As turns out, due to very limited number of parameters, these type of
models are either disfavored by the
current data \cite{Gonzalez-Garcia:2014bfa} or do not generate enough amount of the baryon asymmetry. In order to circumvent this obstacle, in a recent
work \cite{Achelashvili:2016nkr} the setup with two degenerate RHN's and two texture zero $3\tm 2$ Dirac type Yukawa couplings augmented with a
single $\De L=2$ lepton number violating d$=5$ operator has been investigated. All textures, within such setup, giving experimentally viable neutrino mass
matrices have been studied in great details. As turned out \cite{Achelashvili:2016nkr}, some of them together with successful neutrino sector give
interesting predictions and allow to calculate cosmological CP phase in terms of the neutrino CP phase $\de $.

Encouraged by these findings,
in this paper we aim to investigate such construction in details from the viewpoint of the leptogenesis.
Thus, we start our studies with
 the minimal SUSY standard model augmented with two RHNs,
  which at high energy scales
are strictly degenerate in mass. The degeneracy is lifted by the renormalization. As we show,
taking into account the charged lepton Yukawa couplings  into the
 renormalization procedure (where, in a regime of RHN masses$\stackrel{<}{_\sim }\!10^7$~GeV,\footnote{These mass values, we consider within our studies,
  avoid the relic gravitino problem \cite{{Khlopov:1984pf},{Davidson:2002qv}}.}
  the decisive role is played by the tau lepton's Yukawa coupling), the non zero cosmological
 lepton asymmetry emerges at 1-loop level. Moreover, the sufficient baryogenesis
 is realized even with RHN masses near the TeV scale and also with low values of the MSSM parameter $\tan \bt$($\sim 1$).
 As we have mentioned, to make scenario viable, in Ref. \cite{Achelashvili:2016nkr} we have included single $\De L=2$, d$=5$ operator, which
we adopt also in this paper. Inclusion of such terms does not alter RG studies and results mentioned above are robust.
For demonstrative purposes we pick up one of the viable models of \cite{Achelashvili:2016nkr}. That is concrete neutrino texture zero mass matrix
(referred to as the texture $P_1$),
which emerges via integration of two (quasi)
degenerate RHNs and single $\De L=2$, d$=5$ operator. Model's predictive power allows to compute the cosmological
CP phase in terms of  observed neutrino parameters  and CP phases (not measured yet, but predicted by the model).

Note that an approach, similar to the one  we pursue in this paper, could work also within a non SUSY
framework (i.e. within SM augmented with two degenerate RHNs).
However, since for a solution to the gauge hierarchy problem the supersymmetry appears to be a well
motivated (and perhaps the best so far) framework, we choose to perform our investigations within the SUSY setup.

The paper is organized as follows.
  In section  \ref{RG-CP}, we first describe our setup and then, proving emergence of the cosmological CP violation via
 charged lepton Yukawas at 1-loop level, give detailed calculation of
CP violation relevant for the leptogenesis.
In Sect. \ref{nuModel-ResLepGen-UVcompl}
 we present the neutrino scenario (discussed in Ref. \cite{Achelashvili:2016nkr} together with other scenarios),
 with prediction of the CP phase $\de $ and its relation  with the cosmological
CP violation. On this scenario we demonstrate that leptonic asymmetry, induced at quantum level (and computed in Sect. \ref{RG-CP}) leads
 to desirable baryon asymmetry via resonant leptogenesis.
Then we present one example of renormalizable UV completion of our model and prove the robustness of all obtained results.
Appendix \ref{app-RG} includes details and various aspects of the renormalization group (RG) studies.
In appendix \ref{app-scalar-asym} we investigate the effects
of the scalar components of the RHN superfields in the net baryon asymmetry.

\section{Two Quasi-Degenerate RHN and Cosmological CP}
\la{RG-CP}

In this section, we first describe our setup and then give detailed calculation of
CP violation relevant for the leptogenesis.

Our framework is the MSSM augmented with two
right-handed neutrinos $N_{1}$ and $N_{2}$. This extension is enough to build consistent neutrino sector
accommodating the neutrino data \cite{Gonzalez-Garcia:2014bfa} and also to have successful leptogenesis scenario.
The relevant lepton superpotential couplings, we are starting with, are given by:
\begin{equation}
W_{lept}=l^{T}Y_{e}^{\rm diag}e^{c}h_{d}+l^{T}Y_{\nu}Nh_{u}-\frac{1}{2}N^{T}M_{N}N \la{r21},
\end{equation}
where $h_{d}$ and $h_{u}$ are down and up type MSSM Higgs doublet
superfields respectively and  $l^{T}=(l_{1}, l_{2}, l_{3})$, $e^{cT}=(e^{c}_{1}, e^{c}_{2}, e^{c}_{3})$, $N^T=(N_1, N_2)$.
 We  work in a basis in which the charged lepton Yukawa matrix is diagonal and real:
 \beq
 Y_{e}^{\rm diag}={\rm Diag}(\lambda_{e}, \lambda_{\mu}, \lambda_{\tau}).
\eeq
 Moreover, we assume that the RHN mass matrix $M_{N}$ is strictly degenerate at high scale. For the latter we take the GUT scale
 $M_G\simeq 2\cdot 10^{16}$~GeV.\footnote{Degeneracy of $M_N$ can be guaranteed by some symmetry at high energies. For concreteness, we assume
 this energy interval  to be $\geq M_G$ (although the degeneracy at lower energies can be considered as well).}
Therefore, we assume:
 \beq
{\rm at}~~\mu=M_G:~~~M_{N}= \left(\begin{array}{ccc}
 0&1\\
 1&0
\end{array}\right)M(M_G). \la{m01}
\eeq
 This form of $M_{N}$ is crucial for our studies.
  Although it is interesting and worth to study, we do not attempt here to justify the form of $M_N$ (and of the textures
 considered below) by symmetries. Our approach here is rather phenomenological aiming to investigate possibilities,
 outcomes and implications of the textures we consider.
 Since (\ref{m01}) at a tree level leads to the mass degeneracy of the RHN's, it has
interesting implications for resonant leptogenesis
\cite{{Branco:2006hz},{Babu:2007zm},{Babu:2008kp}} and also, as we will see
below, for building predictive neutrino scenarios \cite{Babu:2008kp}, \cite{Achelashvili:2016nkr}.

For the leptogenesis scenario two necessary conditions need to be satisfied. First of all, at the scale $\mu =M_{N_{1,2}}$ the degeneracy between
the masses of $N_1$ and $N_2$ has to be lifted. And, at the same scale, the neutrino Yukawa matrix $\hat Y_{\nu }$ - written in the mass eigenstate  basis of $M_N$, must be such that ${\rm Im}[ (\hat Y_{\nu }^\dag \hat Y_{\nu })_{12}]^2\neq 0$.  [These  can be seen
from Eq. (\ref{res-lept-asym}) with a demand $\ep_{1,2}\neq 0$.]
Below we show that both these are realized by radiative corrections and needed effect already arises at 1-loop level, with
a dominant contribution due to the $Y_e$ Yukawa couplings (in particular from $\lam_{\tau }$) in the RG.

\subsection{Loop Induced Cosmological CP Violation}
\la{sec1}

Radiative corrections are crucial for the cosmological CP violation.
We will start with rediative corrections to the $M_N$ matrix. RG effects cause lifting of the mass degeneracy and, as we will see,
are important also for the phase misalignment (explained below).

At the GUT scale, the $M_N$ has off-diagonal form with $(M_N)_{11}=(M_N)_{22}=0$ [see Eq. (\ref{m01})]. However, at low energies, RG
corrections generate these entries. Thus, we parameterize the matrix $M_N$ at scale $\mu $ as:
\beq
M_{N}(\mu )=\left(
  \begin{array}{cc}
    \de_{N}^{(1)}(\mu ) & 1 \\
    1 & \de_{N}^{(2)}(\mu ) \\
  \end{array}
\right) \!M(\mu ).
\la{MN-with-deltas}
\eeq
While all entries of the matrix $M_N$ run, for our studies will be relevant the ratios $\fr{(M_N)_{11}}{(M_N)_{12}}=\de_{N}^{(1)}$
and $\fr{(M_N)_{22}}{(M_N)_{12}}=\de_{N}^{(2)}$ (for which we will write and solve RG equations below). That's why we have written $M_N$ in a form given
in Eq. (\ref{MN-with-deltas}).
With $|\de_{N}^{(1,2)}|\ll 1$, the $M$ (at scale $\mu =M$) will determine the masses of RHNs $M_1$ and $M_2$, while
$\de_{N}^{(1,2)}$ will be responsible for their splitting and for complexity in $M_N$ (the phase of the overall factor $M$ do not
contribute to the physical CP). As will turn out (see below):
\beq
 \de_{N}^{(1)}=(\de_{N}^{(2)})^*\equiv -\de_N .
 \la{rel-deltN}
 \eeq
 Therefore, $M_N$ is diagonalized by the transformation
$$
U_N^TM_NU_N=M_N^{Diag}={\rm Diag}\l M_1, M_2\r ~,~~~~~{\rm with}~~U_N=P_NO_N{P_N}' ~,
$$
\beq
M_1=|M|\l 1-|\de_N|\r ~,~~~~~~M_2=|M|\l 1+|\de_N|\r ~,
\la{MN-diag-tion}
\eeq
where
$$
P_N={\rm Diag}\l e^{-i\eta/2}, e^{i\eta/2}\r ,~~~
O_N=\fr{1}{\sq{2}}\left(
      \begin{array}{cc}
        1 & -1 \\
        1 & 1 \\
      \end{array}
    \right) ,
~~~{P_N}'={\rm Diag}\l e^{-i\phi_M/2}, ie^{-i\phi_M/2}\r ,
$$
\beq
{\rm with}~~~\eta ={\rm Arg}\l \de_N\r ~,~~~~~\phi_M={\rm Arg}\l M\r .
\la{P-ON-P1}
\eeq

In the $N$'s mass eigenstate basis, the Dirac type neutrino Yukawa  matrix will be $\hat Y_{\nu }=Y_{\nu }U_N$. In the CP asymmetries, the components
$(\hat Y_{\nu }^\dag \hat Y_{\nu })_{21}$ and $(\hat Y_{\nu }^\dag \hat Y_{\nu })_{12}$ appear [see Eq. (\ref{res-lept-asym})].
From (\ref{MN-diag-tion}) and (\ref{P-ON-P1}) we have
\beq
\left [(\hat Y_{\nu }^\dag \hat Y_{\nu })_{21} \right ]^2=- \left [ (O_N^TP_N^*Y_{\nu }^\dag Y_{\nu }P_NO_N)_{21}\right ]^2 ,~~~~
\left [(\hat Y_{\nu }^\dag \hat Y_{\nu })_{12} \right ]^2=- \left [ (O_N^TP_N^*Y_{\nu }^\dag Y_{\nu }P_NO_N)_{12}\right ]^2 .
\la{YdagY-sq}
\eeq
Therefore, we see that the CP violating part should come from the combination $P_N^*Y_{\nu }^\dag Y_{\nu }P_N$, which in a matrix form is:
\beq
P_N^*Y_{\nu }^\dag Y_{\nu }P_N=
\left(
  \begin{array}{cc}
    (Y_{\nu }^\dag Y_{\nu })_{11} & \left |(Y_{\nu }^\dag Y_{\nu })_{12}\right |e^{i(\eta -\eta')}\\
    |(Y_{\nu }^\dag Y_{\nu })_{21}|e^{i(\eta'-\eta)} & (Y_{\nu }^\dag Y_{\nu })_{22} \\
  \end{array}
\right) ,~~~~{\rm with}~~~~\eta'={\rm Arg} [(Y_{\nu }^\dag Y_{\nu })_{21}] ~.
\la{PYYP-form}
\eeq
We see that  $\eta'-\eta$ difference (mismatch) will govern the CP asymmetric decays of the RHNs. Without including the charged lepton Yukawa couplings in the
RG effects we will have $\eta' \simeq \eta $ with a high accuracy. It was shown in Ref.  \cite{Dev:2015wpa} that by ignoring $Y_e$ Yukawas no
CP asymmetry emerges at ${\cal O}(Y_{\nu}^4)$ order and  non zero contributions start only from ${\cal O}(Y_{\nu}^6)$ terms \cite{Pilaftsis:2015bja}.
Such corrections are extremely suppressed for $Y_{\nu } \stackrel{<}{_\sim }1/50$.
Since in our consideration we are interested in cases with $M_{1,2}\stackrel{<}{_\sim }10^7$~GeV giving  $|(Y_{\nu })_{ij}|<7\cdot 10^{-4}$
(well fixed from the neutrino sector and the desired value of the baryon asymmetry), these effects
(i.e. order $\sim Y_{\nu}^6$ corrections) will not have any relevance.
In Ref.  \cite{Babu:2008kp}  in the RG of $M_N$  the effect
of $Y_e$, coming from 2-loop corrections,  was taken into account
and was shown that sufficient CP violation can emerge. Below we show that including $Y_e$  in the  $Y_{\nu}$'s 1-loop RG,
will induce sufficient amount of CP violation. This mainly happens via $\lam_{\tau}$ Yukawa coupling. Thus, below we give detailed investigation of
$\lambda_{\tau}$'s effect.

Using $M_N$'s RG given in Eq. (\ref{MN-2loop-RG}) (of Appendix \ref{app-YM-RGs}), for $\de_N^{(1,2)}$,
which are the ratios $\fr{(M_N)_{11}}{(M_N)_{12}}$
and $\fr{(M_N)_{22}}{(M_N)_{12}}$, [see parametrization in Eq. (\ref{MN-with-deltas})],
we can derive the following RG equations:
$$
16\pi^2 \fr{d}{dt}\de_N^{(1)}\!=\!4(Y_{\nu}^\dag Y_{\nu })_{21}\!+\!2\de_N^{(1)}\! \left [(Y_{\nu}^\dag Y_{\nu })_{11}\!-\!(Y_{\nu}^\dag Y_{\nu })_{22}\right ] \!-\!
2(\de_N^{(1)})^2(Y_{\nu}^\dag Y_{\nu })_{12}\! -\! 2\de_N^{(1)}\de_N^{(2)} (Y_{\nu}^\dag Y_{\nu })_{21}
$$
\beq
-\fr{1}{4\pi^2}(Y_{\nu}^\dag Y_eY_e^\dag Y_{\nu })_{21}+\cdots
\la{deN1-RG}
\eeq
$$
16\pi^2 \fr{d}{dt}\de_N^{(2)}\!=\!4(Y_{\nu}^\dag Y_{\nu })_{12}\!+\!2\de_N^{(2)}\! \left [(Y_{\nu}^\dag Y_{\nu })_{22}\!-\!(Y_{\nu}^\dag Y_{\nu })_{11}\right ] \!-\!
2(\de_N^{(2)})^2(Y_{\nu}^\dag Y_{\nu })_{21}\! -\! 2\de_N^{(1)}\de_N^{(2)} (Y_{\nu}^\dag Y_{\nu })_{12}
$$
\beq
-\fr{1}{4\pi^2}(Y_{\nu}^\dag Y_eY_e^\dag Y_{\nu })_{12}+\cdots
\la{deN2-RG}
\eeq
were in second lines of (\ref{deN1-RG}) and (\ref{deN2-RG}) are given 2-loop corrections depending on $Y_e$. Dots there stand for higher order irrelevant terms. From 2-loop corrections we keep only $Y_e$ dependent terms. Remaining contributions
are not relevant for us.\footnote{Omitted terms  are either strongly suppressed or do not give any significant contribution neither to the CP violation nor to the RHN mass splittings.}
From  (\ref{deN1-RG}) and (\ref{deN2-RG}) we see that dominant contributions come from the first terms of the r.h.s and from those given in the second rows. Other terms give
contributions of order ${\cal O}(Y_{\nu}^4)$ or higher and thus will be ignored. At this approximation we have
\beq
\de_N^{(1)}(t)\simeq \de_N^{(2)*}(t)\equiv -\de_N(t)\simeq
-\fr{1}{4\pi^2} \int_{t}^{t_G}\!\!\!dt ~ \l Y_{\nu }^\dag ({\bf 1}-\fr{1}{16\pi^2}Y_eY_e^\dag )Y_{\nu }\r_{\!21}~
\la{approx-deN12}
\eeq
where $t=\ln \mu $, $t_G=\ln M_G$ and we have used the boundary conditions at the GUT scale  $\de_N^{(1)}(t_G)=\de_N^{(2)}(t_G)=0$.
For evaluation of the integral in (\ref{approx-deN12}) we need to know the scale dependence of $Y_{\nu }$ and $Y_e$. This is
 found in Appendix \ref{app-YM-RGs} by solving the RG equations for  $Y_{\nu }$ and $Y_e$.
Using Eqs. (\ref{approx-Ynu-sol}) and  (\ref{RG-factors}),
the integral of the matrix appearing in (\ref{approx-deN12}) can be written as:
\beq
\int_{t_M}^{t_G} \!\!Y_{\nu }^\dag ({\bf 1}-\fr{1}{16\pi^2}Y_eY_e^\dag ) Y_{\nu } dt \simeq \bar{\ka }(M)Y_{\nu G}^\dag \left(
  \begin{array}{ccc}
    1 & 0 & 0 \\
    0 & 1 & 0 \\
    0 & 0 & \bar r_{\tau }(M) \\
  \end{array}
\right)Y_{\nu G}
\la{int-matrix}
\eeq
where
\beq
 \bar r_{\tau }(M)=\fr{\int_{t_M}^{t_G}\!\! \ka(t)r_{\tau}(t)(1-\fr{\lam_{\tau }^2}{16\pi^2})dt}{\int_{t_M}^{t_G}\! \ka(t)dt}~,~~~~~
 \bar{\ka }(M)=\int_{t_M}^{t_G}\!\! \ka(t)dt~,
\la{bar-r-kapa-2loop}
\eeq
\beq
 r_{\tau }(\mu )=\eta^2_{\tau}(\mu)~,~~~~~\ka(\mu)=\eta^6_t(\mu)\eta^2_{g\nu }(\mu)~
\la{r-kapa}
\eeq
and we have ignored $\lam_{e, \mu }$ Yukawa couplings. For the definition of $\eta $-factors see Eq. (\ref{RG-factors}).
The $Y_{\nu G}$ denote corresponding Yukawa matrix at scale $\mu =M_G$.
On the other hand, we have:
\beq
\left. (Y_{\nu }^\dag Y_{\nu })\right |_{\mu=M}\simeq \ka(M)Y_{\nu G}^\dag \left(
  \begin{array}{ccc}
    1 & 0 & 0 \\
    0 & 1 & 0 \\
    0 & 0 & r_{\tau }(M) \\
  \end{array}
\right)Y_{\nu G}~.
\la{YYnu-M}
\eeq
(Derivations are given in Appendix \ref{app-YM-RGs}.)

Comparing (\ref{int-matrix}) with (\ref{YYnu-M}) we see that difference in these matrix structures (besides overall flavor universal RG factors) are
in the RG factors $r_{\tau }(M)$ and $\bar r_{\tau }(M)$. Without the $\lam_{\tau}$ Yukawa coupling  these factors are equal and there is no mismatch between
the phases $\eta $ and $\eta'$ [defined in Eqs. (\ref{P-ON-P1}) and  (\ref{PYYP-form})] of these matrices. Non zero $\eta' -\eta $ will be due to the deviation, which we parameterize as
\beq
\xi =\fr{\bar r_{\tau }(M)}{r_{\tau }(M)}-1 ~.
\la{xi-shift}
\eeq
This value can be computed numerically by evaluation of the appropriate RG factors. However, it is useful
to have approximate expression for $\xi $, which is given by:
$$
\xi \!\simeq  \!\left [\fr{\lam_{\tau}^2(M)}{16\pi^2}\ln \fr{M_G}{M}
+\fr{1}{3}\fr{\lam_{\tau}^2(M)}{(16\pi^2)^2}
\left [ 3\lam_t^2+6\lam_b^2+10\lam_{\tau}^2-(2c_e^a+c_{\nu}^a)g_a^2\right ]_{\mu=M} \l \!\ln \fr{M_G}{M} \!\r^{\!\!2}\right ]_{\rm 1-loop}
$$
\beq
-~\left [\! \fr{\lam^2_{\tau }(M)}{16\pi^2}\!\right ]_{\rm 2-loop}~,
\la{approx-for-xi}
\eeq
where one and two loop contributions are indicated. Derivation of this expression is given in Appendix \ref{app-YM-RGs}.
As we see, non zero $\xi $ is induced already at 1-loop
level [without 2-loop correction of $\fr{\lam_{\tau }^2}{16\pi^2}$ in Eq. (\ref{bar-r-kapa-2loop})]. However, inclusion of 2-loop correction can contribute to the $\xi $ by amount of $\sim 3-5\%$ (because of $\ln \fr{M_G}{M}$ factor suppression) and we have included it.

Now we are ready to write down quantities which have direct relevance for the leptogenesis. From (\ref{approx-deN12}), with definitions introduced above and by obtained relations, we have:
\beq
|\de_N(M) |e^{i\eta }=\fr{1}{4\pi^2}\fr{\bar \ka (M)}{\ka (M)}\left [ |(Y_{\nu}^\dag Y_{\nu})_{21}|e^{i\eta' } +\xi |(Y_{\nu})_{31}(Y_{\nu})_{32}|e^{i(\phi_{31}-\phi_{32})}\right ]_{\mu=M}
\la{eta-vs-eta1}
\eeq
where $\phi_{31}$ and $\phi_{32}$ are phases of the matrix elements $(Y_{\nu})_{31}$ and $(Y_{\nu})_{32}$ respectively at scale $\mu =M$. Eq. (\ref{eta-vs-eta1}) shows well that
in the limit $\xi \to 0$, we have $\eta =\eta '$, while mismatch of these two phases are due to $\xi \neq 0$. With $\xi \ll 1$, from (\ref{eta-vs-eta1})
we derive:
\beq
\eta -\eta' \simeq \xi \fr{|(Y_{\nu})_{31}(Y_{\nu})_{32}|}{|(Y_{\nu}^\dag Y_{\nu})_{21}|} \sin (\phi_{31}-\phi_{32}-\eta')~.
\la{eta-eta1-aprox}
\eeq
We stress, that the 1-loop renormalization of the $Y_{\nu }$ matrix plays the leading role in generation of $\xi$, i.e. in the
CP violation. [This is also demonstrated by Eq. (\ref{approx-for-xi}).] 

The value of $|\de_N(M) |$, which characterizes the mass splitting between the RHN's, can be computed taking absolute value of both sides of
(\ref{eta-vs-eta1}):
\beq
|\de_N(M) |=\fr{\ka_N}{4\pi^2}\left |(Y_{\nu}^\dag Y_{\nu})_{21} +\xi (Y_{\nu})_{31}(Y_{\nu}^*)_{32} \right |_{\mu=M}\ln \fr{M_G}{M}~,~~~~~{\rm with}~~~~~~
\ka_N=\fr{\bar{\ka}(M)}{\ka (M)\ln \fr{M_G}{M}}~.
\la{abs-deltaN}
\eeq
These expressions can be used upon the calculation of the leptogenesis, which we will do in the next section for one concrete
model of the neutrino mass matrix.

\section{Predictive Neutrino Texture and Baryon Asymmetry}
\la{nuModel-ResLepGen-UVcompl}

In this section we apply obtained results within the setup of the couplings (\ref{r21}) augmented by single
$\De L=2$, d$=5$ operator. As was shown in \cite{Achelashvili:2016nkr}, this could lead to the successful and predictive neutrino
sectors. With the addition of this  d$=5$ operator, the results obtained above can remain intact.
We consider one neutrino scenario which allows to predict the CP phase $\de $ and relate it with the cosmological
CP violation leading to desirable baryon asymmetry via resonant leptogenesis.
First we discuss the neutrino sector and then turn to the investigation of the leptogenesis.
At the end, we present one possible renormalizable UV completion (giving rise to $\De L=2$, d$=5$ operator which we utilize) maintaining all  obtained results.

\subsection{$P_1$ Neutrino Texture: Relating $\de $ and Cosmological CP}
\la{sect-P1-texture}

In the work of Ref.  \cite{Achelashvili:2016nkr}, within the setup of two (quasi) degenerate RHNs was studied neutrino mass
matrices emerged from two zero $3\times2$ Yukawa textures in combination of one $d=5$ entry. In this way, all experimentally
viable neutrino mass matrices have been investigated, which also predicted CP violation and gave promise for successful leptogenesis.
Here, for concreteness we consider one scenario of the neutrino mass matrix - called in \cite{Achelashvili:2016nkr} the $P_1$ type texture -
and show that it admits having calculable CP violation.

Thus,  we consider the Yukawa matrix with the form:
\beq
Y_{\nu}=
\begin{pmatrix}
0 & 0\\
a_{2}e^{i\alpha_{2}} & b_{2}e^{i\beta_{2}}\\
a_{3}e^{i\alpha_{3}} &  b_{3}e^{i\beta_{3}}

\end{pmatrix}
=
\begin{pmatrix}
e^{ix} & 0&0\\
0 & e^{iy}&0\\
0 & 0&e^{iz}
\end{pmatrix}
\begin{pmatrix}
0 & 0\\
a_{2} & b_{2}\\
a_{3} &b_{3}e^{i\phi}
\end{pmatrix}
\begin{pmatrix}
e^{i\omega} & 0\\
0 & e^{i\rho}
\end{pmatrix}, \la{t4212}
\eeq
{\rm with}
\beq
\omega=\alpha_{2}-\beta_{2}+\rho,\quad y= \beta_{2}-\rho,\quad z=
\alpha_{3}-\alpha_{2}+\beta_{2}-\rho, \quad \phi=\alpha_{2}-\alpha_{3}+\beta_{3}-\beta_{2}.
\la{t42121}
\eeq
where,  only one phase $\phi$ will be relevant for the cosmological CP asymmetry. The phases $x,y,z$ can be removed by proper phase
redefinitions of the states $l_i$ and $e^c_i$.
Using this and the form of $M_N$, given in Eq. (\ref{m01}), via see-saw
formula we get the following contribution to the neutrino mass matrix:
\beq
M^{ss}_{\nu}=-\langle
h^{0}_{u}\rangle^{2} Y_{\nu}M^{-1}_{N}Y^{T}_{\nu}. \la{seesaw}
\eeq
Besides this, we include the $\mathrm{d}=5$ operator
\beq
\frac{\tilde d_{5}
e^{ix_{5}}}{M_*}l_1l_2h_uh_u~,
\la{d5-op}
\eeq
 where $M_*$ and $\tilde d_{5}$ are some cut off scale and dimensionless coupling respectively. With proper phase redefinitions of
 $l_i$ states, without loss of any generality, both these can be taken real and the phase $x_5$ selected as $x_5=\om +\rho -{\rm arg}(M)$.
  The origin of the operator (\ref{d5-op}) and
consistency of our construction will be discussed in Sect. \ref{sec-consist}.
Taking into account these and Eq. (\ref{seesaw}), the neutrino mass matrix at scale $M$ will have the
form:
\beq
M_{\nu }(M)=-
\begin{pmatrix}
0 & d_{5}&0\\
d_{5}&2a_{2}b_{2}&a_{3}b_{2}+a_{2}b_{3}e^{i\phi}\\
0&a_{3}b_{2}+a_{2}b_{3}e^{i\phi}&2a_{3}b_{3}e^{i\phi}
\end{pmatrix}\frac{v_u^2(M)}{M\cdot e^{-i(\omega+\rho)}},~~~~{\rm with}~~~ d_5=\tl{d}_5\fr{|M|}{M_*}~,
\la{t124}
\eeq
where in $M_N$ we have ignored $(1,1)$ and $(2,2)$ elements, which are induced at 1-loop level and are so suppressed that have no impact
on light neutrino masses and mixings.
By the renormalization (discussed in Appendix \ref{app-nuRG}) for the neutrino mass matrix at scale
$M_Z$ we obtain:
\beq
M_{\nu}(M_Z)=\begin{pmatrix}
0&d_5&0\\
d_5&2a_2b_2&(a_3b_2+a_2b_3e^{i\phi})r_{\nu3}\\
0&(a_3b_2+a_2b_3e^{i\phi})r_{\nu3}&2a_3b_3e^{i\phi}r^2_{\nu3}
\end{pmatrix}{\bar{m}},~~~~{\rm with}~~~  \bar m=-\frac{r_{\bar{m}}v_{u}^2(M_Z)}{M\cdot e^{-i(\omega+\rho)}} ,
\la{M}
\eeq
where the couplings $a_i, b_i, d_5$ and phases appearing in (\ref{M}) are given at scale $M$.
The RG factors $r_{\nu 3}$ and $r_{\bar m}$ are given in Eqs. (\ref{r-nu3}) and  (\ref{r-mbar}) respectively.
The neutrino mass matrix (\ref{M}) is of the $P_1$ type investigated in details in \cite{Achelashvili:2016nkr}.

Noting that we are working in a basis of diagonal charged lepton mass matrix, the neutrino mass matrix can be related to the lepton mixing matrix $U$ by:
\beq
M_{\nu}=PU^{*}P^{'}M_{\nu}^{\rm diag}U^{+}P \la{nu1}
\eeq
where  $M_{\nu}^{\rm diag}=(m_{1},m_{2},m_{3})$
and the phase matrices and $U$ are:
\beq
 P={\rm
Diag}(e^{i\omega_{1}},e^{i\omega_{2}},e^{i\omega_{3}}),\quad
P^{'}={\rm Diag}(1,e^{i\rho_{1}},e^{i\rho_{2}})\la{nu2}
\eeq \beq
U= \left(\begin{array}{ccc}
c_{13}c_{12} &c_{13}s_{12}&s_{13}e^{-i\delta}\\
-c_{23}s_{12}-s_{23}s_{13}c_{12}e^{i\delta}&
c_{23}c_{12}-s_{23}s_{13}s_{12}e^{i\delta}&s_{23}c_{13}\\
s_{23}s_{12}-c_{23}s_{13}c_{12}e^{i\delta}&-s_{23}c_{12}-c_{23}s_{13}s_{12}e^{i\delta}&c_{23}c_{13}
\end{array}\right)\la{nu3}
\eeq

%
%
\begin{table}
\vs{-0.8cm}
{\small
 $$\begin{array}{|c|c|c|c|}
\hline
\vs{-0.35cm}
 & & & \\
\delta & \rho_{1}& \rho_{2}&{\rm works~ with} \\
\hline

 &&& {\rm NH}, ~\sin^{2}\theta_{23}=0.49~{\rm  and ~best~ fit~ values} \\
 \pm 0.378  & \pm 3.036 &\pm 2.696 & {\rm    [of ~Eq.~ (\ref{nu-inputs})]~for~ remaining ~oscillation~ parameters ,}\\
 &&& (m_{1},m_{2},m_{3})=(0.00613,0.0106,0.0499), m_{\beta\beta}=0 \\
\hline
\end{array}$$
\vs{-0.5cm}
}
\caption{Results from $P_{1}$ type texture of Eq. (\ref{M}). Masses are given in eVs.}
 \label{tab1}
\end{table}
%
%
%

As was discussed in details in \cite{Achelashvili:2016nkr}, the
texture (\ref{M}) allows only normal neutrino mass hierarchy. Using the conditions $M_{\nu}^{(1,1)}=M_{\nu}^{(1,3)}=0$ in Eq. (\ref{nu1}),
 we obtain the following predictions:
\beq
m^{2}_{3}=\frac{\Delta m^{2}_{atm}+\Delta
m^{2}_{sol}c^{2}_{12}}{1-s^{2}_{13}\cot^{2}_{23}(1+t^{2}_{13})^{2}-t^{4}_{13}},~~~~
\cos\rho_{1}=\frac{m^{2}_{3}t^{4}_{13}-m^{2}_{1}c^{4}_{12}-m^{2}_{2}s^{4}_{12}}{2m_{1}m_{2}c^{2}_{12}s^{2}_{12}},
\la{h}
\eeq
 \beqs
\delta=\arg[m_{1}c^{2}_{12}+m_{2}s^{2}_{12}e^{i\rho_{1}}]-\arg[m_{1}-m_{2}e^{i\rho_{1}}],
\eeqs \beq
\rho_{2}=\pm
\pi-\arg[m_{1}c^{2}_{12}+m_{2}s^{2}_{12}e^{i\rho_{1}}]+2\arg[m_{1}-m_{2}e^{i\rho_{1}}] ,
\la{m}
\eeq
where, by definition $\De m^{2}_{\mathrm{atm}}=m^{2}_{3}-m^{2}_{2}$ and $\De m^{2}_{\mathrm{sol}}=m^{2}_{2}-m^{2}_{1}$.
With the inputs
$$
\sin^{2}\theta_{12}=0.304, \quad \sin^{2}\theta_{23}=0.49, \quad \sin^{2}\theta_{13}=0.0218,
$$
\beq
 \De m^{2}_{\mathrm{atm}}=0.002382~{\rm eV}^2, ~~~~~
 \De m^{2}_{\mathrm{sol}}=7.5\cdot10^{-5}~{\rm eV}^2,
\la{nu-inputs}
\eeq
we obtain the values:
$$
m_{1}=0.00613~{\rm eV},\quad m_{2}=0.0106~{\rm eV}, \quad m_{3}=0.0499~{\rm eV},
$$
\beq
\rho_1=\pm3.036,\quad \delta=\pm0.378, \quad \rho_2=\pm 2.696.
\eeq
Notice that besides $\sin^{2}\theta_{23}$ all inputs of Eq. (\ref{nu-inputs}) are taken to be the best fit values \cite{Gonzalez-Garcia:2014bfa}.
 The results are summarized in Table \ref{tab1}.

At the same time, from (\ref{nu1}) we  have the relations:
\beq
2a_{2}b_{2}\bar m=e^{2i\omega_{2}}\mathcal{A}_{22},\quad
2a_{3}b_{3}e^{i\phi}\bar mr^{2}_{\nu 3}=e^{2i\omega_{3}}\mathcal{A}_{33}, \quad
(a_{3}b_{2}+a_{2}b_{3}e^{i\phi})\bar mr_{\nu 3}=e^{i(\omega_{2}+\omega_{3})}\mathcal{A}_{23}, \la{q0}
\eeq
with
\beq
\mathcal{A}_{ij}=U^{\ast}_{i1}U^{\ast}_{j1}m_{1}+U^{\ast}_{i2}U^{\ast}_{j2}m_{2}e^{i\rho_{1}}+U^{\ast}_{i3}U^{\ast}_{j3}m_{3}e^{i\rho_{2}}.
\la{qq}
\eeq
Note, that from the neutrino sector all $\mathcal{A}_{ij}$ numbers are determined with the help of
zero entries in matrix of Eq. (\ref{M}).
With the help of the phases appearing in (\ref{t4212}), without loss of generality we can take $a_i, b_i>0$. With this, from the
equations of (\ref{q0}) we can express $|\bar m|$ and the couplings $a_3, b_{2,3}$ in terms of $a_2$ and $|M|$ as follows:
\beq
|\bar m|\!=\!\fr{v_u^2(M_Z)}{|M|}r_{\bar m},~~
a_3\!=\!\fr{a_2}{r_{\nu 3}}\left |\fr{1}{{\cal A}_{22}}\l \!{\cal A}_{23}\pm \sqrt{ {\cal A}_{23}^2\!-\!{\cal A}_{22}{\cal A}_{33}} \r \right |,~~
b_2\!=\!\fr{1}{a_2}\fr{|{\cal A}_{22}|}{2|\bar m|}~,~~b_3\!=\!\fr{1}{a_3}\fr{|{\cal A}_{33}|}{2|\bar m|r_{\nu 3}^2}~.
\la{rels-ab}
\eeq
Also, for the phase $\phi $ we get the following prediction:
\beq
\phi ={\rm Arg}\left[\left(\frac{\mathcal{A}_{23}}{\sqrt{\mathcal{A}_{22}\mathcal{A}_{33}}}\mp
\sqrt{\frac{\mathcal{A}^{2}_{23}}{\mathcal{A}_{22}\mathcal{A}_{33}}-1}\right)^{2}\right].
\la{pred-phi}
\eeq
Notice, that there is a pair of solutions. When  for the $a_3$'s expression in Eq. (\ref{rels-ab}) we are taking the `$+$' sign,
in Eq. (\ref{pred-phi}) we should take the sign   `$-$', and vice versa.

From these, using results given in Table \ref{tab1}, we find numerical value of $\phi$:
$$
{\rm for}~~~\de=+0.378:~~~~~~~\phi_{+}=+1.287~,~~~~~\phi_{-}=-1.287~,
$$
\beq
{\rm for}~~~\de=-0.378:~~~~~~~\phi_{+}=-1.287~,~~~~~\phi_{-}=+1.287~,
\la{value-phi}
\eeq
where $\phi $'s subscripts correspond to the signs taken in (\ref{pred-phi}).
These and the relations of (\ref{rels-ab}) will be used upon calculation of the baryon asymmetry, which we do in the next subsection.

\subsection{Resonant Leptogenesis}

The CP asymmetries $\ep_1$ and $\ep_2$
generated by out-of-equilibrium decays of the quasi-degenerate fermionic components of $N_1$ and $N_2$ states respectively are given by
\cite{Pilaftsis:1997jf, Pilaftsis:2003gt}:\footnote{In appendix \ref{app-scalar-asym} we investigate the contribution to the baryon asymmetry
via decays of the scalar components of the RHN superfields. As we show, these effects are less than $3\%$.}
\beq
\ep_1=\fr{{\rm Im}[(\hat{Y}_{\nu }^{\dagger}\hat{Y}_{\nu })_{21}]^2}
{(\hat{Y}_{\nu }^{\dagger}\hat{Y}_{\nu })_{11}(\hat{Y}_{\nu }^{\dagger}\hat{Y}_{\nu })_{22}}
\fr{\l M_2^2-M_1^2\r M_1\Ga_2}{\l M_2^2-M_1^2\r^2+M_1^2\Ga_2^2}~,
~~~~~~~~~~~~\ep_2=\ep_1(1\lrar 2)~.
\la{res-lept-asym}
\eeq
Here $M_1, M_2$ (with $M_2>M_1$) are the mass eigenvalues of the RHN  mass matrix. These masses, within our scenario, are given in
(\ref{MN-diag-tion}) with the splitting parameter given in Eq. (\ref{abs-deltaN}).
The decay widths of fermionic RHN's are given by $\Ga_i=\fr{M_i}{4\pi}(\hat Y_{\nu}^{\dag}\hat Y_{\nu})_{ii}$.
 Moreover, the imaginary part of $[(\hat{Y}_{\nu }^{\dagger}\hat{Y}_{\nu })_{21}]^2$
will be computed with help of (\ref{YdagY-sq}) and (\ref{PYYP-form}) with the relevant phase given in Eq. (\ref{eta-eta1-aprox}).
Using  general expressions (\ref{eta-eta1-aprox}) and (\ref{abs-deltaN}) for the neutrino model discussed in previous subsection, we get:
\beq
\eta \!-\!\eta' \simeq -\xi \fr{\fr{a_2b_2}{a_3b_3}\sin \phi }{(\fr{a_2b_2}{a_3b_3}\!+\!\cos \phi )^2\!+\!\sin^2\phi }~,~~~~~~
|\de_N(M) |\!=\!\fr{\ka_N}{4\pi^2}\left |a_2b_2\!+\!a_3b_3(1\!+\!\xi )e^{i\phi} \right |\ln \fr{M_G}{M}~.
\la{eta-eta1-abs-deltaN-model}
\eeq
With these, since we know the possible values of the phase $\phi $ [see Eq. (\ref{value-phi})], and with the help of the relations (\ref{rels-ab}) we can
compute $\ep_{1,2}$ in terms of $|M|$  and $a_2$.
Recalling that the  lepton asymmetry is converted to the baryon asymmetry via sphaleron processes
\cite{Kuzmin:1985mm}, with the relation $\frac{n_b}{s}\simeq -1.48\times10^{-3}({\kappa_f}^{(1)}\epsilon_1+{\kappa_f}^{(2)}\epsilon_2) $
we can compute the baryon asymmetry.
For the efficiency factors ${\kappa_f}^{(1,2)}$  we will use the extrapolating expressions  \cite{Giudice:2003jh} (see Eq. (40) in Ref. \cite{Giudice:2003jh}),
with  ${\kappa_f}^{(1)}$ and ${\kappa_f}^{(2)}$ depending on the mass scales
${\tilde{m}}_1=\frac{v_u^2(M)}{M_1}(\hat{Y}_{\nu}^{\dag}{\hat{Y_{\nu}}})_{11}$ and ${\tilde{m}}_2=\frac{v_u^2(M)}{M_2}(\hat{Y}_{\nu}^{\dag}{\hat{Y_{\nu}}})_{22}$ respectively.

%
%
\begin{table}
\vs{-0.8cm}
{\small
 $$\begin{array}{|c|c|c|c|c|}
\hline
\vs{-0.35cm}
&&&&\\
 & {\rm Case}~ {\bf (I_{-})}  & {\rm Case}~ {\bf (I)}  & {\rm Case}~ {\bf (II_{-})} &{\rm Case}~ {\bf (II)}  \\
\vs{-0.35cm}
&&&&\\
\hline
\vs{-0.35cm}
&&&&\\
 m_t(m_t) & 162.77~{\rm GeV} & 163.48 ~{\rm GeV}&  162.77 ~{\rm GeV}& 163.48 ~{\rm GeV}\\
\vs{-0.35cm}
&&&&\\
\hline
\vs{-0.4cm}
&&&&\\
 M_S &10^3~{\rm GeV} &10^3~{\rm GeV} &2\cdot 10^3 ~{\rm GeV}&2\cdot 10^3 ~{\rm GeV}\\
\hline
\end{array}$$
\vs{-0.5cm}
}
\caption{Cases with different values of $m_t(m_t)$ and $M_S$.}
 \label{tab-cases}
\end{table}
%

%
%
%
%
%
\begin{table}
\vs{0.3cm}

 $$\begin{array}{|c|c|c|c|c|c|c|c|c|}
\hline
{\rm Case}  & {\rm M(GeV}) & \tan \bt  &  r_{\nu 3} & r_{\bar m} & r_{v_u} & \ka_N &
 10^{5}\!\tm \! \xi  & 10^{11}\!\tm \!\l \!\fr{n_b}{s}\!\r_{\rm max} \\
\hline
{\bf (I_{-})} & 3\cdot 10^3 & 1.63 & \simeq 1 &  0.8861  & 0.9713 & 1.230 & 5.678 & 8.573 \\
\hline
 {\bf (I.1)} &  3\cdot 10^3 & 1.636 & \simeq 1 & 0.8849 & 0.9709 & 1.242 & 5.729 & 8.565 \\
 \hline
 {\bf (I.2)} &  10^4 & 1.665 & \simeq 1 & 0.8343 &0.953  & 1.211 & 5.490 & 8.564  \\
\hline
 {\bf (I.3)} &  10^5 & 1.72 &\simeq 1 & 0.7530 & 0.9218 & 1.1596 &5.0317  & 8.559 \\
 \hline
 {\bf (I.4)} &  10^6 & 1.775 & \simeq 1 & 0.6883 & 0.8944 & 1.118  & 4.574 & 8.557 \\
 \hline
 {\bf (I.5)} &  10^7 & 1.831 & \simeq 1 & 0.6369 &0.8703  & 1.0834 &4.118  &  8.565\\
 \hline
 \hline
 {\bf (II_{-})} &6\cdot 10^3 & 1.608 & \simeq 1 & 0.8685& 0.9677 & 1.197 & 5.462 & 8.557\\
 \hline
 {\bf (II.1)} &6\cdot 10^3 &1.615 & \simeq 1& 0.8670& 0.9673 & 1.206 & 5.515& 8.564\\
 \hline
 {\bf (II.2)} & 10^4 & 1.627 &\simeq 1  &0.8468 & 0.9600&  1.195 &5.416 &8.563\\
 \hline
 {\bf (II.3)} & 10^5 &1.681 & \simeq 1 &0.7671 &0.9295 & 1.147&4.968  & 8.557 \\
 \hline
 {\bf (II.4)} & 10^6 & 1.736 & \simeq 1 &0.7034  & 0.9027 &1.108  & 4.523 &8.565\\
 \hline
 {\bf (II.5)} & 10^7 & 1.79 & \simeq 1 & 0.6524  & 0.8790 & 1.076 & 4.072& 8.564\\
 \hline
\end{array}$$
\vs{-0.5cm}
\caption{Baryon asymmetry for various values of $M$ and for minimal (allowed) value of $\tan \bt $.
The values of $\l \fr{n_b}{s}\!\r_{\rm max}$ given here are obtained for all cases of Eq.  (\ref{value-phi}), but
for different values of $a_i, b_j$.
(For phase sign choices see (\ref{pred-phi}), (\ref{value-phi}) and comments after these Eqs.)  }
 \vs{-0.3cm}
 \label{results-for-P1}
\end{table}
%
%
%

Within our studies we will consider the RHN masses $\simeq |M|\stackrel{<}{_\sim }10^7$~GeV. With this, we will not have the relic gravitino
problem \cite{{Khlopov:1984pf},{Davidson:2002qv}}.
For the simplicity, we consider all  SUSY particle masses to be equal to  $M_S< |M|$, with $M_S$ identified with the SUSY scale, below
which we have just SM.
As it turns out, via the RG factors, the asymmetry
  also depends on the top quark mass. Therefore, we will consider cases given in Table \ref{tab-cases}, were cases of low top quark masses by  $1$-$\si $
deviation are included [i.e cases  ${\bf (I_{-})}$ and  ${\bf (II_{-})}$]. It is remarkable that the observed baryon asymmetry
\beq
\l \! \fr{n_b}{s}\!\r_{\rm exp} =\l 8.65\pm 0.085\r \tm 10^{-11}
\la{nbs-exp}
\eeq
(the recent value reported according to WMAP and Planck  \cite{Ade:2015xua}), can be obtained even for low values of the MSSM parameter $\tan \bt =\fr{v_u}{v_d}$
(defined at the SUSY scale $\mu = M_S$). This, for different cases and different values of $M$, is demonstrated in Table \ref{results-for-P1}.
For the calculations we have used the RG factors found by numerical computations. The details of this procedure, appropriate boundary and
matching conditions are given in Appendix \ref{app-baund-match}.

\begin{figure}
\begin{center}
\leavevmode
\leavevmode
\includegraphics{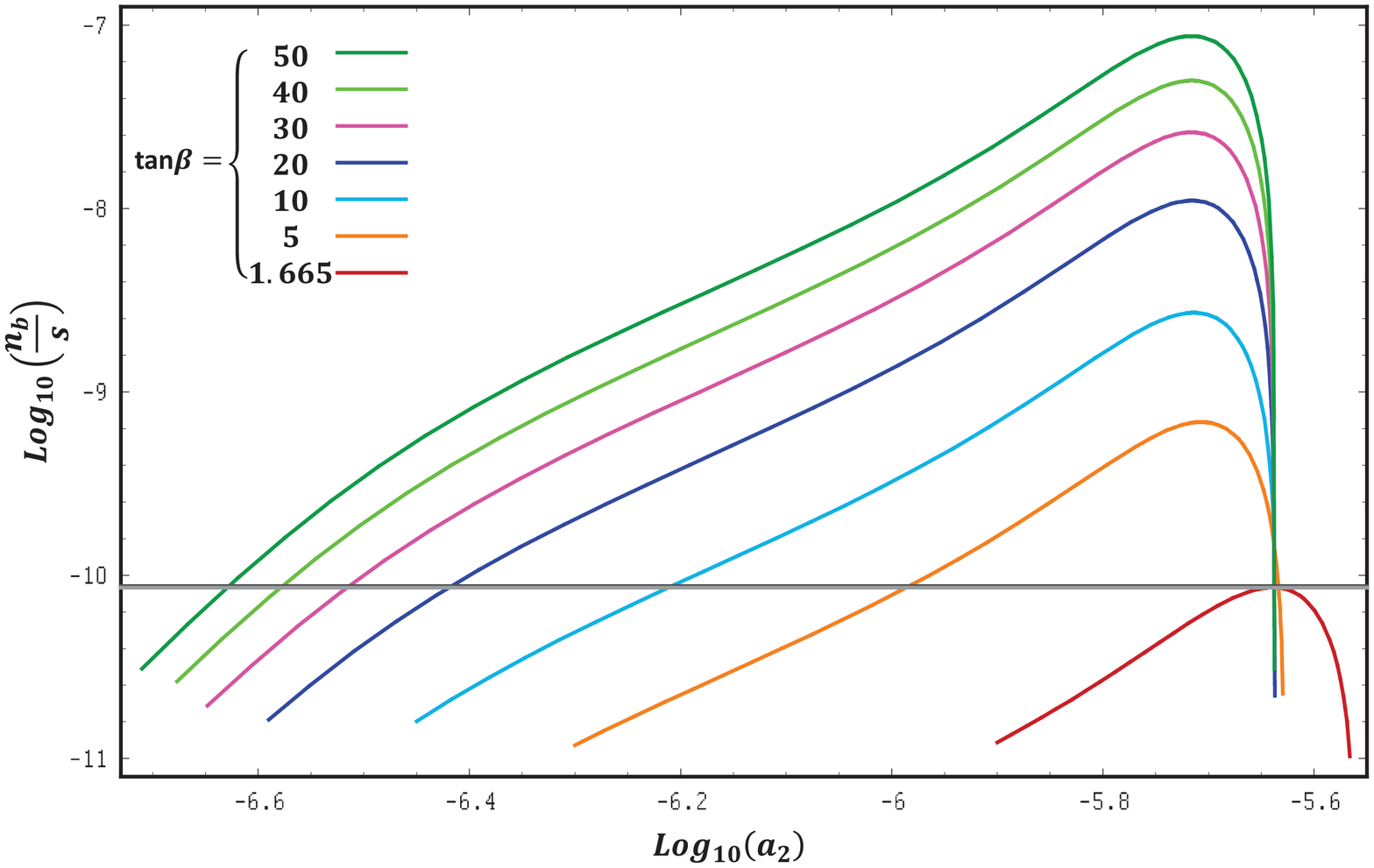}  
     \includegraphics{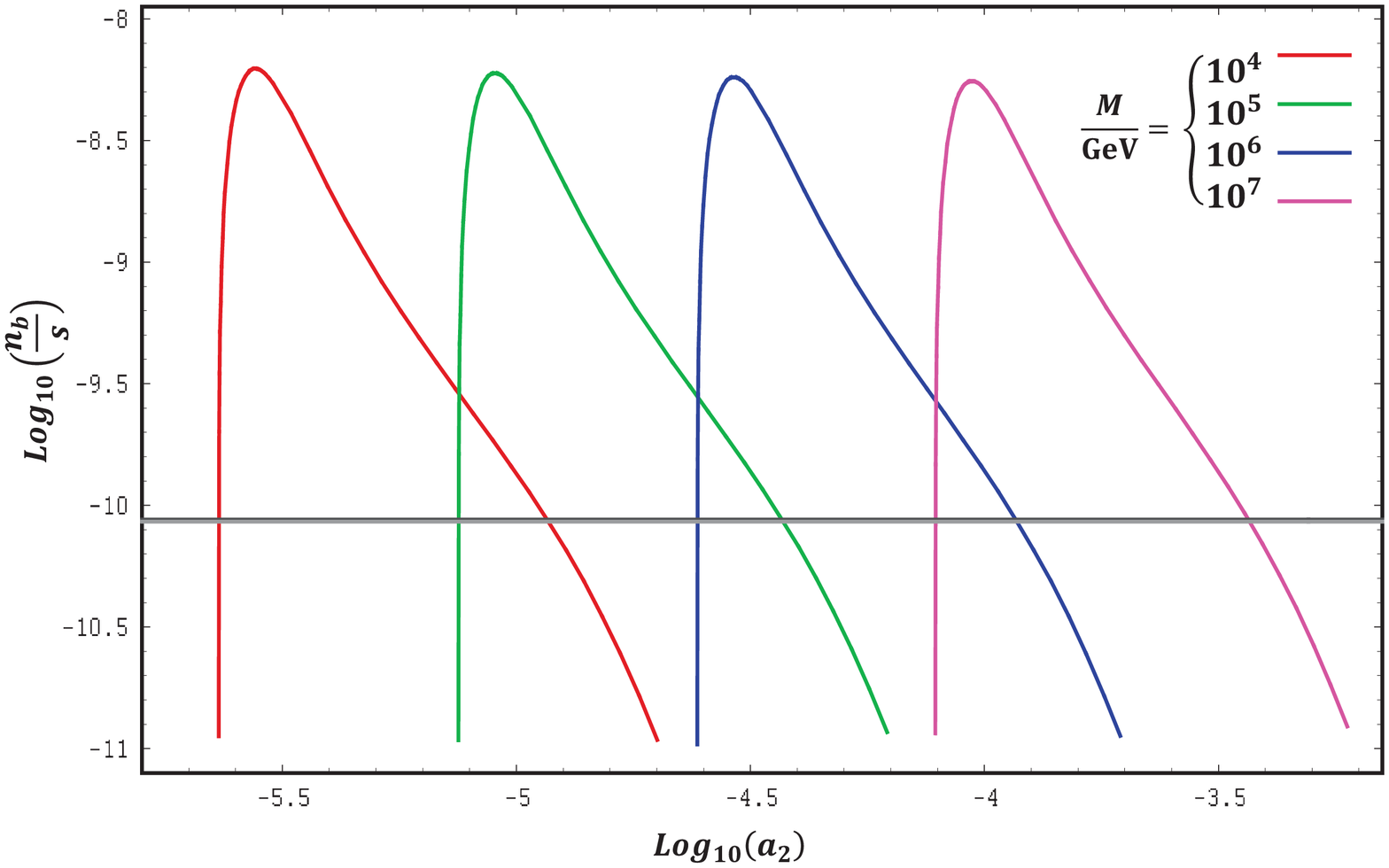}
\end{center}
\vs{5cm}
\caption{{\bf Left:} Curves for case {\bf (I)} [see Table \ref{tab-cases}], with $M=10^4$~GeV, $\de =+0.378$, $\phi=\phi_{+}=+1.287$ and with different
values of $\tan \bt $.
{\bf Right:} Curves for case {\bf (II)} [see Table \ref{tab-cases}], with $\tan \bt=15$~GeV, $\de =-0.378$, $\phi=\phi_{+}=-1.287$ and with different
values of $M$.
Gray horizontal  bands correspond to the experimental value of the baryon asymmetry within the 1-$\si $ range
given in Eq. (\ref{nbs-exp}).}
\label{fig1-2}
\end{figure}

While Table   \ref{results-for-P1} deals with cases of the low $\tan \bt $, in plots of Figure \ref{fig1-2} we
show baryon asymmetries as  functions of $a_2$ (the logs of these values for convenience) for different values of the parameters $M_S, M, \tan \bt $ and the phases $\phi $ of Eq. (\ref{value-phi}).
We see that needed baryon asymmetry is obtained for a wide range of phenomenologically interesting values of  parameters.
With the values of $a_2$ giving the needed values of the baryon asymmetry, we have also calculated [via relations of Eq. (\ref{rels-ab})]
the values of $a_3, b_{2,3}$, which also turned out to be suppressed, i.e. $a_3, b_{2,3} \stackrel{<}{_\sim }a_2$.

%
%
%

\subsection{Renormalizable UV Completion and Consistency Check}
\la{sec-consist}

Upon building the neutrino mass matrix (\ref{t124}), together with see-saw contribution (\ref{seesaw}) (emerged via integration of $N_{1,2}$ states)
we have used the $d=5$ operator (\ref{d5-op}). Here we present one renormalizable completion of the model, which gives the latter operator.
Also we check  the whole construction and show what conditions should be satisfied in order
to have fully consistent model without affecting obtained results.

For building fully renormalizable model, we introduce two additional RHN states
 ${\cal N}$ and $\bar{\cal N}$ with the following superpotential couplings:
\beq
\lam l_1{\cal N}h_u+\bar{\lam }l_2\bar{\cal N}h_u-M_*{\cal N}\bar{\cal N}~.
\la{l1-l2-calN}
\eeq
 With these and the couplings of (\ref{r21})-(\ref{m01}),  (\ref{t4212}), after removing the phases $x,y,x, \om, \rho$ in $Y_{\nu}$
 (by proper redefinition of the fields) without loss of generality $\bar{\lam }$ and $M_*$ can be taken real and ${\rm arg}(\lam)={\rm arg}(\bar m)$.
 Thus, the full (i.e. 'extended') Yukawa and RHN matrices will be:
\begin{equation}
\begin{array}{ccc}
 & {\begin{array}{cccc}
 N_1 & \!N_2  &\!\!{\cal N} & \!\!\bar{\cal N}
\end{array}}\\ \vspace{2mm}
Y_{\nu}^{ext}=\begin{array}{c}
l_1 \\ l_2 \\l_3
 \end{array}\!\!\!\!\! &{\left(\begin{array}{cccc}
0 &0 & \lam & 0 \\
a_2 & b_2 & 0 & \bar{\lam } \\
a_3 & b_3e^{i\phi } & 0 & 0 \\
\end{array}\right) }~,
\end{array}  \!\!  ~~~~~
\begin{array}{ccc}
 & {\begin{array}{cccc}
\hs{-0.3cm} N_1 & N_2  &~{\cal N} &~~ \bar{\cal N}
\end{array}}\\ \vspace{2mm}
M_N^{ext}=\begin{array}{c}
N_1 \\ N_2 \\{\cal N} \\ \bar{\cal N}
 \end{array}\!\!\!\!\! &{\left(\begin{array}{cccc}
0 & M & 0 & 0 \\
                  M & 0 & 0 & 0 \\
                  0 & 0 & 0 & M_* \\
                  0 & 0 & M_* & 0 \\
\end{array}\right) }~.
\end{array}  \!\!  ~~~~~
\label{ext-matr}
\eeq
With these forms, integration of heavy RHN states leads to the neutrino mass matrix
\beq
M_{\nu }=-v_u^2Y_{\nu}^{ext}(M_N^{ext})^{-1}(Y_{\nu}^{ext})^T,
\la{nu-matrix-UV}
\eeq
which, as desired, indeed has the form of (\ref{t124}) with
\beq
d_5=|\lam |\bar{\lam }\fr{|M|}{M_*}~.
\la{d5-derivation}
\eeq

Furthermore, one should make sure that via loops the couplings $\lam $ and $\bar{\lam } $
instead of zeros in the textures of Eq. (\ref{ext-matr}) do not induce entries which would  affect and/or spoil  the results of
the neutrino sector and leptogenesis. To check this, one can apply  1-loop RGs for the neutrino Yukawas and
RHN masses. Namely, in Eqs. (\ref{Ynu-RG}) and (\ref{MN-2loop-RG}) with the replacements $Y_{\nu }\to Y_{\nu}^{ext}$,
$M_N\to M_N^{ext}$ we can estimate the 1-loop contributions due to the $\lam , \bar{\lam }$
couplings.\footnote{Since (as we have seen) the couplings $a_i, b_i$ are small, their corrections in the RG of $Y_{\nu}^{ext}$ do not harm anything.}
Since the structure of $Y_{\nu}^{ext}$ may be altered only by the second term at r.h.s of (\ref{Ynu-RG}), we will calculate only contribution
due to this type of entry. By the same reason, for the $M_N^{ext}$'s correction, we will focus only on the first term (and on it's transpose)
 at the r.h.s of Eq. (\ref{MN-2loop-RG}).
Doing so,  with an assumption $M_*> |M|$, at scale $\mu=M_* $ we
obtain:
$$
\begin{array}{ccc}
\de Y_{\nu}^{ext}\approx \!-\fr{3}{16\pi^2}\!
 {
 \left(\begin{array}{cccc}
0 &0 & \lam |\lam |^2& 0 \\
a_2\bar{\lam }^2 & b_2\bar{\lam }^2 & 0 & \bar{\lam }(a_2^2+b_2^2+\bar{\lam }^2) \\
\tm  & \tm  & 0 &  \bar{\lam }(a_2a_3+b_2b_3e^{i\phi })\\
\end{array}\right)\!\!\ln \fr{M_G}{M_*} }~,~~
\end{array}
$$
\begin{equation}
\begin{array}{ccc}
\de M_N^{ext}\approx  \!-\fr{1}{8\pi^2}\!
 {\left(\begin{array}{cccc}
\tm & \tm & a_2\bar{\lam}M_* & b_2\bar{\lam}M \\
                  \tm & \tm & b_2\bar{\lam}M_* & a_2\bar{\lam}M \\
                  a_2\bar{\lam}M_* & b_2\bar{\lam}M_* & 0 & (|\lam |^2+\bar{\lam}^2)M_* \\
                  b_2\bar{\lam}M  & a_2\bar{\lam}M & (|\lam |^2+\bar{\lam}^2)M_* & 0 \\
\end{array}\right)\!\!\ln \fr{M_G}{M_*}  }~,
\end{array}  \!\!  ~~~~~
\label{1-loop-cor-ext-matr}
\eeq
where we have taken into account that at  scale $\mu =M_G$ the couplings $Y_{\nu}^{ext}$, $M_N^{ext}$ have forms given in Eq. (\ref{ext-matr}).
In (\ref{1-loop-cor-ext-matr}) `$\tm $' stand for the corrections which do not depend on $\lam $ and/or $\bar{\lam }$.
Comparing (\ref{1-loop-cor-ext-matr}) with (\ref{ext-matr}) we see that the structure of $Y_{\nu}^{ext}$ is not changed and $\de Y_{\nu}^{ext}$
can be negligible for $\lam, \bar{\lam}\stackrel{<}{_\sim }\lam_{\tau}/10$. In fact, from the neutrino sector, we have
\beq
d_5|\bar m|=|{\cal A}_{12}|\simeq 1.07\cdot 10^{-11}~{\rm GeV} .
\la{d5-from-neutr}
\eeq
[see Eqs. (\ref{M}) and (\ref{qq}) for definitions.]
With this, on the other hand, we have
\beq
d_5\approx 4.15\cdot 10^{-12}\l \fr{M}{10^4{\rm GeV}}\r \l \fr{1}{\sin \bt }\r^2\l \fr{0.85}{r_{\bar m}}\r~.
\la{d5-numbers}
\eeq
With this and $M_*= (3-10)M$, the (\ref{d5-derivation}) can be satisfied by the selection
\beq
|\lam |\approx \bar{\lam}=\l d_5\fr{M_*}{|M|}\r^{1/2}
\simeq (3.5-6.4)\cdot 10^{-6}\l \!\fr{M}{10^4{\rm GeV}}\!\r^{\!\!1/2} \!\!\!\l \fr{1}{\sin \bt }\r \!\!\l \fr{0.85}{r_{\bar m}}\r^{\!\!1/2}~.
\la{lamb-vals}
\eeq
This in turn gives:
\beq
{\rm for}~~M\stackrel{<}{_\sim }10^{7}~{\rm GeV},~~~\tan \bt >1.6~~ \Longrightarrow   ~~|\lam |\approx \bar{\lam} <3\cdot 10^{-4}~.
\la{bound-lam}
\eeq
We checked and made  sure that, for such small values of $\lam, \bar{\lam}$, the corrections $\de Y_{\nu}^{ext}$ and  $\de M_{N}^{ext}$
are affecting neither the neutrino sector, nor the leptogenesis. We have also checked that 2-loop corrections are
very suppressed too and can be safely ignored.
The  selection $M_*= (3-10)M$ is convenient because the states ${\cal N}$, $\bar{\cal N}$ (having the mass $M_*$)
decouple  earlier than the states $N_{1,2}$  and will not contribute to the leptogenesis process.
With all these we conclude that the results obtained in previous subsections stay robust.

Closing this section, we comment (as was also noted in Sect. 2), that throughout our studies we have not
 attempted to explain and justify texture zeros by symmetries. Our approach here was to consider such textures
 which give predictive and consistent scenario allowing to calculate cosmological CP violation.
 The forms of the matrices in Eqs.  (\ref{m01}), (\ref{t4212}) and/or (\ref{ext-matr})  with specific coupling selections are such that their
 structures and model's predictive power (as was demonstrated) are not ruined by radiative corrections. For our purposes
 this was already satisfactory. More fundamental explanation should be pursued elsewhere.

\subsubsection*{Acknowledgments}
We thank A. Pilaftsis for discussions and helpful comments.
The work is partially supported by Shota Rustaveli National Science Foundation (Contract No. DI/12/6-200/13).
 Z.T. thanks CETUP* (Center for Theoretical Underground Physics and Related Areas)
 for its hospitality and partial support during 2016 Summer Program.
 Z.T. also would like to express a special thanks to the Mainz Institute for Theoretical Physics (MITP) for its hospitality and support.

\appendix

\renewcommand{\theequation}{A.\arabic{equation}}\setcounter{equation}{0}

\section{Renormalization Group Studies}
\la{app-RG}

\subsection{Running of $Y_{\nu }, Y_e$ and $M_N$ Matrices and Approximation for $\xi $}
\la{app-YM-RGs}

RG equations for the charged lepton and neutrino Dirac Yukawa matrices, appearing in the superpotential of Eq. (\ref{r21}), at 1-loop order have the forms
\cite{Martin:1993zk}, \cite{Antusch:2002ek}:
\beq
16\pi^2 \fr{d}{dt}Y_e=3Y_eY_e^\dag Y_e+Y_{\nu}Y_{\nu}^\dag Y_e+
Y_e \left [ {\rm tr}\l 3Y_d^\dag Y_d+Y_e^\dag Y_e \r
 -c_e^ag_a^2\right ]~,~~~~~c_e^a=(\fr{9}{5}, 3, 0) ,
\la{Ye-RG}
\eeq
\beq
16\pi^2 \fr{d}{dt}Y_{\nu}=Y_eY_e^\dag Y_{\nu}+3Y_{\nu}Y_{\nu}^\dag Y_{\nu}+
Y_{\nu} \left [ {\rm tr}\l 3Y_u^\dag Y_u+Y_{\nu}^\dag Y_{\nu} \r
 -c_{\nu }^ag_a^2\right ]~,~~~~~c_{\nu }^a=(\fr{3}{5}, 3, 0) .
\la{Ynu-RG}
\eeq
$g_a=\l g_1, g_2, g_3\r $ denote gauge couplings of $U(1)_Y, SU(2)_w$ and $SU(3)_c$ gauge groups respectively. Their 1-loop RG have forms
$16\pi^2 \fr{d}{dt}g_a=b_ag_a^3$, with $b_a=(\fr{33}{5}, 1, -3)$, where the hypercharge of $U(1)_Y$ is taken in $SU(5)$ normalization.

The RG for the RHN mass matrix at 2-loop level has the form \cite{Antusch:2002ek}:
$$
16\pi^2\fr{d}{dt}M_N=2M_NY_{\nu}^{\dag}Y_{\nu}
-\fr{1}{8\pi^2}M_N\left [ Y_{\nu}^{\dag}Y_eY_e^{\dag}Y_{\nu}+Y_{\nu}^{\dag}Y_{\nu}Y_{\nu}^{\dag}Y_{\nu}+
Y_{\nu}^{\dag}Y_{\nu}{\tr}(3Y_u^\dag Y_u+Y_{\nu}^{\dag}Y_{\nu})\right ]
$$
\beq
+\fr{1}{8\pi^2}M_NY_{\nu}^{\dag}Y_{\nu}\l \fr{3}{5}g_1^2+3g_2^2\r +({\rm transpose})~,
\la{MN-2loop-RG}
\eeq

Let's start with  renormalization of the $Y_{\nu }$'s matrix elements.
Ignoring in Eq. (\ref{Ynu-RG}) the ${\cal O}(Y_{\nu}^3)$ order entries (which are very small because within our studies
$|(Y_{\nu})_{ij}|\stackrel{<}{_\sim }10^{-4}$), and from charged fermion Yukawas keeping $\lam_{\tau}$ and $\lam_t$, we will have:
\beq
16\pi^2 \fr{d}{dt}\ln (Y_{\nu})_{ij}\simeq \de_{i3}\lam_{\tau }^2+3\lam_t^2-c_{\nu }^ag_a^2~.
\la{Ynu-RG-approx}
\eeq
This gives the solution
\beq
(Y_{\nu})_{ij}(\mu)=(Y_{\nu G})_{ij}(\eta_{\tau}(\mu))^{\de_{i3}}\eta_t^3(\mu)\eta_{g\nu }(\mu ) ,
\la{approx-Ynu-sol}
\eeq
where $Y_{\nu G}$ denotes Yukawa matrix at scale $M_G$ and the scale dependent RG factors are given by:
$$
\eta_{t, b, \tau}(\mu)\!=\!\exp \!\l \!\!-\fr{1}{16\pi^2}\!\!\int_t^{t_G}\!\!\lam^2_{t, b, \tau } (t')dt' \!\r ,~~
\eta_a(\mu )\!=\!\exp \!\l \!\!\fr{1}{16\pi^2}\!\!\int_t^{t_G}\!\!\! g^2_a (t')dt' \!\r
$$
\beq
\eta_{g\nu }(\mu )\!=\exp \l \!\!\fr{1}{16\pi^2}\!\int_t^{t_G}\!\!\!\!c_{\nu}^ag^2_a (t')dt' \!\!\r
=\eta_1^{3/5} (\mu ) \eta_2^3 (\mu ),~~~
{\rm with}~~~t=\ln \mu ~,~t'=\ln \mu' ~,~~t_G=\ln M_G .
\la{RG-factors}
\eeq
From these, for the combination $Y_{\nu }^\dag Y_{\nu }$ at scale $\mu =M$ we get expression given in Eq. (\ref{YYnu-M}).

On the other hand, for the RHN mass splitting and for the phase mismatch [depending on $\xi $ defined in Eq. (\ref{xi-shift})], the  integrals/factors of Eqs. (\ref{int-matrix}), (\ref{bar-r-kapa-2loop}),  (\ref{r-kapa})
and (\ref{YYnu-M}) will be relevant.
For obtaining approximate analytical results [for the expression of $\fr{\bar r_{\tau }(M)}{r_{\tau }(M)}$] we will use expansions.
Namely, we introduce the notation
\beq
{\cal K}=\ka r_{\tau}\!\l 1-\fr{\lam^2_{\tau }}{16\pi^2}\r
\la{K}
\eeq
and make a Taylor expansion of ${\cal K}(t)$  and $\ka(t)$ near the point $t=t_M$, in powers of $(t-t_M)$. As will turn out,
this will allow to calculate $\xi=\fr{\bar r_{\tau }(M)}{r_{\tau }(M)}-1$  in  powers of $\fr{\lam_{\tau}^2}{16\pi^2}$ [and possibly
in powers of other couplings appearing in higher degrees - together with appropriate  $\fr{1}{16 \pi^2}$ factors].
We have:
$$
{\cal K}(t)={\cal K}(t_M)+{\cal K}'(t_M)(t-t_M)+\fr{1}{2}{\cal K}''(t_M)(t-t_M)^2+\cdots
$$
\beq
\ka (t)=\ka (t_M)+\ka'(t_M)(t-t_M)+\fr{1}{2}\ka''(t_M)(t-t_M)^2+\cdots
\la{exand-K-kapa}
\eeq
where primes denote derivatives with respect to $t$. Plugging  these in Eq. (\ref{bar-r-kapa-2loop}) and performing integration we will get:
$$
\bar r_{\tau }(M)\!=
 \fr{{\cal K}(t_M)}{\ka (t_M)}\l  \!1+\!\fr{1}{2}\fr{{\cal K}'(t_M)}{{\cal K}(t_M)}(t_G\!-\!t_M)\!+
\!\fr{1}{6}\fr{{\cal K}''(t_M)}{{\cal K}(t_M)}(t_G\!-\!t_M)^2\!+\!\cdots \r \! \tm
$$
\beq
\tm \l  \!1+\!\fr{1}{2}\fr{\ka '(t_M)}{\ka (t_M)}(t_G\!-\!t_M) \!+\!\fr{1}{6}\fr{\ka ''(t_M)}{\ka (t_M)}(t_G\!-\!t_M)^2 \!+\!\cdots \r^{\!-1}\!\!\!\!\!.
\la{expand-bar-r}
\eeq
Using in (\ref{expand-bar-r}) expression (\ref{K}) for ${\cal K}$ and keeping in expansion terms up to the  $(t-t_M)^2$, we get
\beq
\fr{\bar r_{\tau}(M)}{r_{\tau}(M)} -1
\simeq \fr{1}{2}\left. \fr{r_{\tau }'}{r_{\tau }}\right |_{t=t_M}\!\!\!\!(t_G-t_M)
+\fr{1}{6}\l \fr{r_{\tau}''}{r_{\tau}}+\fr{1}{2}\fr{\ka' r_{\tau}'}{\ka r_{\tau}}\r_{t=t_M} \!\!\!(t_G-t_M)^2 -\fr{\lam^2_{\tau }(M)}{16\pi^2}~.
\la{expand-r-log2}
\eeq
As we see, the  flavor universal RG factor $\ka $ drops out at first order of $(t_G-t_M)$.
Last term in Eq. (\ref{expand-r-log2}) is due to the 2-loop correction
in the RG of $M_N$ [in particular $M_NY_{\nu}^\dag Y_eY_e^\dag Y_{\nu}$ term of r.h.s of Eq. (\ref{MN-2loop-RG})].
Remaining terms are due to 1-loop corrections, proving that cosmological CP violation emerges already at 1-loop level.

Using in (\ref{expand-r-log2}) expressions for the scale factors given in
 Eqs. (\ref{RG-factors}), (\ref{r-kapa}), the RG for $\lam_{\tau}$ [easily obtained from Eq. (\ref{Ye-RG})]  and keeping terms up to the order of
$\fr{1}{(16\pi^2)^2}$, we obtain the expression for $\xi $ given in Eq. (\ref{approx-for-xi}).

\subsection{Neutrino Mass Matrix Renormalization}
\la{app-nuRG}

In the energy interval $M_S\leq \mu <M$ (where $M_S$ is the SUSY scale)
the RG for the neutrino mass matrix is \cite{Chankowski:1993tx}, \cite{Antusch:2002ek}:
\beq
M_S\leq \mu <M :~~~~~16\pi^2\fr{d}{dt}M_{\nu }=Y_eY_e^\dag M_{\nu }+M_{\nu }Y_e^*Y_e^T+M_{\nu }\left [ 6{\rm tr}\l Y_u^\dag Y_u\r -2c_{\nu}^ag_a^2\right ] .
\la{Mnu-RG-SUSY}
\eeq
Below  the $M_S$ scale, effectively we have SM and the RG is \cite{Chankowski:1993tx}:
\beq
\mu<M_S:~~~~~16\pi^2\fr{d}{dt}M_{\nu }=\fr{1}{2}Y_eY_e^\dag M_{\nu }+\fr{1}{2}M_{\nu }Y_e^*Y_e^T+
M_{\nu }\left [ {\rm tr}\l 6Y_u^\dag Y_u \!+\!6Y_d^\dag Y_d\!+\!2Y_e^\dag Y_e\r-3g_2^2+4\lam \right ]~,
\la{Mnu-RG-2HDM}
\eeq
where $\lam $ is the SM Higgs self-coupling (emerging from the self-interaction term $\lam (H^\dag H)^2$ of the SM Higgs doublet $H$).
We will also need the RG evaluation of the VEVs $v_u$ and $v$, which in appropriate energy intervals are given by
\cite{{Grimus:1979pj},{Pendleton:1980as},{Arason:1991ic},{Das:2000uk}}:
\beq
\mu>M_S:~~~~~~~~16\pi^2\fr{d}{dt}v_u=v_u\l -3\lam_t^2+\fr{1}{4}c_{\nu}^ag_a^2 \r ,~
\la{RG-vu}
\eeq
\beq
\mu<M_S:~~~~~16\pi^2\fr{d}{dt}v=v\l -3\lam_t^2-3\lam_b^2-\lam_{\tau }^2+\fr{3}{4}c_{\nu}^ag_a^2 \r .
\la{RG-v}
\eeq
At scale $M$, after decoupling of the RHN states, the neutrino mass matrix is formed with the form:
\beq
M_{\nu }^{ij}(M)=
\left(
  \begin{array}{ccc}
    \tm & \tm & \tm \\
    \tm & \tm & \tm \\
    \tm & \tm & \tm \\
  \end{array}
\right)\fr{v_u^2(M)}{M} ~,
\la{Mnu-M}
\eeq
where `$\tm $' stand for entries depending on Yukawa couplings. After renormalization, keeping $\lam_{\tau}, \lam_t$ and $g_a$ in the above RGs,
 for the neutrino mass matrix at scale $M_Z$ we obtain:
\beq
M_{\nu }^{ij}(M_Z)=
\left(
  \begin{array}{ccc}
    \tm & \tm & (\tm )\!\cdot \!r_{\nu 3}\\
    \tm & \tm & (\tm )\!\cdot \! r_{\nu 3}\\
    (\tm )\!\cdot \! r_{\nu 3}& (\tm )\!\cdot \! r_{\nu 3} & (\tm )\!\cdot \! r_{\nu 3}^2\\
  \end{array}
\right)\bar m ~,~~~~{\rm with}~~~~~\bar m=\fr{v^2(M_Z)s^2_{\bt }}{M}r_{\bar m}~,
\la{Mnu-MZ}
\eeq
where `$\tm $' denotes entries determined at scale $M$  corresponding to those in (\ref{Mnu-M}), and
RG factors are given by
\beq
r_{\nu 3}\!=\!\l \!\fr{\eta_{\tau }(t_Z)}{\eta_{\tau }(t_{M_S})}\!\r^{1/2}
\!\l \!\fr{\eta_{\tau }(t_{M_S})}{\eta_{\tau }(t_{M})}\!\r ~,
\la{r-nu3}
\eeq

\beq
r_{\bar m}\!=\eta_{\lam }^4\!\l \!\fr{\eta_t(t_{m_t})}{\eta_t(t_M)}\!\r^{12} \!\!\l \!\fr{\eta_b(t_Z)}{\eta_b(t_{M_S})}\!\r^{12}
\!\!\l \!\fr{\eta_{\tau }(t_Z)}{\eta_{\tau }(t_{M_S})}\!\r^{4}
\!\!\l \!\fr{\eta_2(t_Z)}{\eta_2(t_M)}\!\r^{\!\fr{15}{2}}
\!\!\l \!\fr{\eta_1^{3/5}(t_Z)\eta_1^{2/5}(t_{M_S})}{\eta_1(t_M)}\!\r^{\!\fr{3}{2}}~,
\la{r-mbar}
\eeq
where
\beq
\eta_{\lam }=\exp \l -\fr{1}{16\pi^2}\int_{t_{m_h}}^{t_{M_S}}\lam (t)dt \r ~,
\la{eta-lam}
\eeq
and remaining $\eta $-factors are defined in Eq. (\ref{RG-factors}).

We will also need the RG factor relating the VEV $v_u(M)$ to  the $v(M_Z)$. Using Eqs. (\ref{RG-vu}) and  (\ref{RG-v}) we obtain:
\beq
r_{v_u}\!=\!\fr{v_u(M)}{v(M_Z)s_{\bt }}\!=\!\l \!\fr{\eta_t(t_{m_t})}{\eta_t(t_M)}\!\r^3
\!\!\l \!\fr{\eta_b(t_Z)}{\eta_b(t_{M_S})}\!\r^{3}
\!\!\l \!\fr{\eta_{\tau }(t_Z)}{\eta_{\tau }(t_{M_S})}\!\r
\!\!\l \!\fr{\eta_2^3(t_Z)\eta_2^{-2}(t_{M_S})}{\eta_2(t_M)}\!\r^{\!\fr{3}{4}}
\!\!\l \!\fr{\eta_1^3(t_Z)\eta_1^{-2}(t_{M_S})}{\eta_1(t_M)}\!\r^{\!\fr{3}{20}}~.
\la{r-vu}
\eeq

\subsection{Boundary and Matching Conditions }
\la{app-baund-match}

For finding the RG factors, appearing in the baryon asymmetry, we  numerically solve renormalization group equations from
the scale $M_Z$ up to the $M_G\simeq 2\cdot 10^{16}$~GeV scale. For simplicity, for all SUSY particle masses we take common mass scale $M_S$.
Thus, in the energy interval $M_Z\leq \mu <M_S$, the Standard Model RGs for  $\ov{\rm MS}$ coupling constants are used.
However, in the interval  $M_S\leq \mu \leq M_G$, since we are dealing with the SUSY, the RGs for the $\ov{\rm DR}$ couplings are applied.
Below we give boundary and matching conditions for the gauge couplings $g_{1,2,3}$, for Yukawa constant $\lam_{t,b,\tau}$ and for the Higgs
self-coupling $\lam $.

\vs{0.2cm}
\hs{-0.5cm}{\bf Gauge couplings}

We choose our inputs for the $\ov{\rm MS}$ gauge couplings at scale $M_Z$ as follows:
$$
\al_1^{-1}(M_Z)=\fr{3}{5}c_w^2\al_{em}^{-1}(M_Z)+\fr{3}{5}c_w^2\fr{8}{9\pi }\ln \fr{m_t}{M_Z}~,~~~~~
\al_2^{-1}(M_Z)=s_w^2\al_{em}^{-1}(M_Z)+s_w^2\fr{8}{9\pi }\ln \fr{m_t}{M_Z}~,
$$
\beq
\al_3^{-1}(M_Z)=\al_s^{-1}(M_Z)+\fr{1}{3\pi }\ln \fr{m_t}{M_Z}~,
\la{in-alpMZ}
\eeq
where logarithmic terms $\ln \fr{m_t}{M_Z}$ are due to the top quark threshold correction \cite{Hall:1980kf}, \cite{Arason:1991ic}.
Taking $\al_s(M_Z)=0.1185$, $\al_{em}^{-1}(M_Z)=127.934$ and $s_w^2=0.2313$, from (\ref{in-alpMZ}) we obtain:
$$
\al_1^{-1}(M_Z)=59.0057+\fr{8c_w^2}{15\pi }\ln \fr{m_t}{M_Z}~,~~~
\al_2^{-1}(M_Z)=29.5911+\fr{8s_w^2}{9\pi }\ln \fr{m_t}{M_Z}~,
$$
\beq
\al_3^{-1}(M_Z)=8.4388+\fr{1}{3\pi }\ln \fr{m_t}{M_Z}~.
\la{num-in-alpMZ}
\eeq
With these inputs we run $g_{1,2,3}$ via the 2-loop RGs from $M_Z$ up to the scale $M_S$.

At scale $\mu =M_S$ we use the matching conditions between $\ov{\rm DR}-\ov{\rm MS}$ gauge couplings \cite{{Antoniadis:1982vr},{Martin:1993yx}}:
\beq
{\rm at}~\mu =M_S:~~~
\fr{1}{\al_1^{\rm \ov{DR}}} = \fr{1}{\al_1^{\rm \ov{MS}}}~,~~~~\fr{1}{\al_2^{\rm \ov{DR}}} = \fr{1}{\al_2^{\rm \ov{MS}}}-\fr{1}{6\pi }~,~~~~
\fr{1}{\al_3^{\rm \ov{DR}}} = \fr{1}{\al_3^{\rm \ov{MS}}}-\fr{1}{4\pi }~.
\la{SUSY-alpha-DR-MS}
\eeq
Above the scale
$M_S$ we apply 2-loop SUSY RG equations in  $\ov{\rm DR}$ scheme \cite{Martin:1993zk}.

\vs{0.2cm}
\hs{-0.5cm}{\bf Yukawa Couplings and $\lam $}

At the scale $M_S$ all SUSY states decouple and we are left with the Standard Model with one Higgs doublet.
Thus, the third family Yukawa couplings and the self-coupling are determined as:
$$
\lam_t(m_t)=\fr{m_t(m_t)}{v(m_t)}~,~~~~\lam_b(M_Z)=\fr{2.89{\rm GeV}}{v(M_Z)}~,~~~~\lam_{\tau}(M_Z)=\fr{1.746{\rm GeV}}{v(M_Z)}~,~~~
$$
\beq
\lam(m_h)=\fr{1}{4}\l \fr{m_h}{v(m_h)}\r^2 ,~~~~~{\rm with}~~~v(M_Z)=174.1~{\rm GeV}~,~~~m_h=125.15~{\rm GeV}~,
\la{yuk-MZ-mt}
\eeq
where $m_t(m_t)$ is the top quark running mass related to the pole mass as:
\beq
m_t(m_t)=p_tM_t^{pole}~.
\la{run-pole-mt}
\eeq
The factor $p_t$ is $p_t\simeq 1/1.0603$ \cite{pole-run}, while the recent measured value of the top's pole mass is \cite{ATLAS:2014wva}:
\beq
M_t^{pole}=(173.34\pm 0.76)~{\rm GeV}.
\la{exp-top-pole}
\eeq
We take the values of (\ref{yuk-MZ-mt}) as boundary conditions for solving 2-loop RG equations \cite{Machacek:1983fi}, \cite{Arason:1991ic}
for $\lam_{t, b, \tau}$ and $\lam $ from the $M_Z$ scale up to the scale $M_S$.

Above the $M_S$ scale, we have MSSM states including two doublets
$h_u$ and $h_d$, which couple with up type quarks and down type quarks and charged leptons respectively.
Thus, the third family Yukawa couplings at $M_S$ are $\approx \lam_t(M_S)/s_{\bt } ,  \lam_b(M_S)/c_{\bt }$
and $\lam_{\tau }(M_S)/c_{\bt }$, with $s_{\bt }\equiv \sin \bt, c_{\bt }\equiv \cos \bt $. Above the scale
$M_S$ we apply 2-loop SUSY RG equations in  $\ov{\rm DR}$ scheme \cite{Martin:1993zk}. Thus, at $\mu =M_S$ we use the matching conditions
between $\ov{\rm DR}-\ov{\rm MS}$ couplings:
$$
{\rm at}~\mu =M_S:~~~~~~~~~~~~~~\lam_t^{\rm \ov{DR}}\simeq \fr{\lam_t^{\rm \ov{MS}}}{s_{\bt }}
\left [ 1+\fr{1}{16\pi^2}\l \fr{g_1^2}{120}+\fr{3g_2^2}{8}-\fr{4g_3^2}{3}\r \right ] ,
$$
\beq
\lam_b^{\rm \ov{DR}}\simeq \fr{\lam_b^{\rm \ov{MS}}}{c_{\bt }}
\left [1\!+\!\fr{1}{16\pi^2}\l \fr{13g_1^2}{120}+\fr{3g_2^2}{8}-\fr{4g_3^2}{3}\r \right ],~~~~
\lam_{\tau }^{\rm \ov{DR}}\simeq \fr{\lam_{\tau }^{\rm \ov{MS}}}{c_{\bt }}
\left [1\!+\!\fr{1}{16\pi^2}\l -\fr{9g_1^2}{40}+\fr{3g_2^2}{8}\r \right ] ,
\la{SUSYyuk-DR-MS}
\eeq
where expressions in brackets of r.h.s of the relations are due to the $\ov{\rm DR}-\ov{\rm MS}$ conversions \cite{Martin:1993yx}.
 With Eq. (\ref{SUSYyuk-DR-MS})'s matchings  we run corresponding couplings from the scale $M_S$ up to the $M_G$ scale.
 Throughout  the paper, above the mass scale $M_S$ without using
  the superscript $\ov{\rm DR}$ we assume the couplings determined in this scheme.

\renewcommand{\theequation}{B.\arabic{equation}}\setcounter{equation}{0}

\section{Contribution to the Baryon Asymmetry from $\tl N$ Decays}
\la{app-scalar-asym}

Impact of the decays of the right handed sneutrinos - the scalar partners of the RHNs -
was estimated in \cite{Babu:2008kp}  for specific textures. Here we give more detailed investigation
and give results for the neutrino model discussed in Sect. \ref{sect-P1-texture}.

We will need to derive masses of the RH sneutrinos and their couplings to the components of the  superfields $l$ and $h_u$.
For this purpose, we should include the soft breaking terms
\beq
V_{SB}^{\nu}=\tl l^TA_{\nu}\tl Nh_u-\fr{1}{2}\tl N^TB_N\tl N+{\rm h.c.}+\tl l^{\dag}m^2_{\tl l}\tl l+\tl N^{\dag}m^2_{\tl N}\tl N~,
\la{V-SB-nu}
\eeq
which, together with the superpotential couplings, will be relevant. As it turns out,
relevant will be $A_{\nu}$ and $B_N$ couplings. Therefore, first we will study their renormalization. After this, we investigate
masses of the physical RH sneutrinos and their couplings to the lepton superfield components. These, at the end, will be
used for the calculation of the contribution in the baryon asymmetry via the RH sneutrino decay processes.

\subsection{Renormalization of Soft $A_{\nu}$ and $B_N$ Terms}
\la{soft}

From general expressions of Ref. \cite{Martin:1993zk} we can derive RGs for $A_{\nu}$ and $B_N$, which at 1-loop level have the forms:
$$
16\pi^2\fr{d}{dt}A_{\nu }=Y_eY_e^\dag A_{\nu }+2\hat A_eY_e^\dag Y_{\nu }+5Y_{\nu }Y_{\nu }^\dag A_{\nu }\!+\!
A_{\nu }\!\left [{\rm tr}(3Y_u^\dag Y_u+Y_{\nu }^\dag Y_{\nu })+4Y_{\nu }^\dag Y_{\nu }-c_{\nu}^ag_a^2\right ]
$$
\beq
+2Y_{\nu }\!\left [{\rm tr}(3Y_u^\dag \hat A_u+Y_{\nu }^\dag A_{\nu })+c_{\nu}^ag_a^2M_{\tl V_a}\right ]~,
\la{RG-Anu}
\eeq
\beq
16\pi^2\fr{d}{dt}B_N=2B_NY_{\nu }^\dag Y_{\nu }+2Y_{\nu }^T Y_{\nu }^*B_N+4M_NY_{\nu }^\dag A_{\nu }+4A_{\nu }^T Y_{\nu }^*M_N~.
\la{RG-B-N}
\eeq
Note that,
applying these expressions for the third generation states we can get expressions of  \cite{Baer:2001vw} (see Eqs. (17) and (55) of this Ref., which
uses slightly different definitions for the couplings).
These results are also compatible with those given in \cite{Ibarra:2008uv} (with replacements $Y\to Y^T$, $A\to A^T$).

We parameterize the matrix $B_N$ as
\beq
B_N=(M_N)_{12}m_B\left(
  \begin{array}{cc}
    \de_{BN}^{(1)} & 1 \\
    1 & \de_{BN}^{(2)} \\
  \end{array}
\right),
\la{BN-matrix}
\eeq
where all entries $(M_N)_{12}, m_B$, $\de_{BN}^{(1,2)}$ run and their RGs can be derived from the RG equations given above.
For the matrix $A_{\nu }$, let's use the parametrization
\beq
A_{\nu }=m_Aa_{\nu }~,
\la{A-nu-matr}
\eeq
 where $m_A$ is a constant and the elements of the $a_{\nu }$
matrix run.
The matrix $\hat A_e$ is
\beq
\hat A_e={\rm Diag}\l A_e, A_{\mu }, A_{\tau }\r
\la{Ae-matr}
\eeq
(similar to the structure of $Y_e$ Yukawa matrix).
We will use the following boundary conditions:
$$
{\rm at}~\mu=M_G:~~~~a_{\nu }=Y_{\nu}~,~~~~\de_{BN}^{(1)}=\de_{BN}^{(2)}=0 ,~~~\hat A_e=m_A{\rm Diag }\l \lam_e, \lam_{\mu },\lam_{\tau }\r
$$
\beq
\hat A_u=m_AY_{uG}~,~~~~\hat A_d=m_AY_{dG}~,
\la{b-conds}
\eeq
which assume proportionality (alignment) of the soft SUSY breaking terms with the corresponding superpotential couplings.

With (\ref{BN-matrix}), (\ref{A-nu-matr}), using (\ref{RG-B-N}) we have:
\beq
16\pi^2\fr{d}{dt}\de_{BN}^{(1)}\simeq 4(Y_{\nu }^\dag Y_{\nu })_{21}+8\fr{m_A}{m_B}(Y_{\nu }^\dag a_{\nu })_{21} ~,~~~~
16\pi^2\fr{d}{dt}\de_{BN}^{(2)}\simeq 4(Y_{\nu }^\dag Y_{\nu })_{12}+8\fr{m_A}{m_B}(Y_{\nu }^\dag a_{\nu })_{12}.
\la{RG-de-BN}
\eeq
Due to RG effects, the alignment between $Y_{\nu}$ and $a_{\nu }$ (which holds at the GUT scale) is violated.
In particular:
\beq
16\pi^2\fr{d}{dt}\l \fr{(a_{\nu })_{ij}}{(Y_{\nu })_{ij}}\r \simeq 2\de_{i3}\fr{\lam_{\tau}A_{\tau}}{m_A}+
\fr{2}{m_A}(3\lam_tA_t+c_{\nu }^ag_a^2M_{\tl V_a})~,
\la{RG-a-Y}
\eeq
where at r.h.s. we kept third family couplings, gauge couplings and gaugino masses.
From this we derive
$$
a_{\nu }\simeq
\left(
  \begin{array}{ccc}
    1+\ep_0 & 0 & 0 \\
    0 & 1+\ep_0 & 0 \\
    0 & 0 & 1+\ep_0+\ep \\
  \end{array}
\right)\!Y_{\nu }
$$
\beq
{\rm with}~~~\ep_0= -\fr{1}{8\pi^2m_A}\int_{t}^{t_G} \!\!\!\!dt(3\lam_tA_t+c_{\nu }^ag_a^2M_{\tl V_a})~,~~~~~~
\ep =-\fr{1}{8\pi^2m_A}\int_{t}^{t_G} \!\!\!\!dt\lam_{\tau}A_{\tau }~.
\la{sol-a-nu}
\eeq
Using (\ref{sol-a-nu}) in Eqs. (\ref{RG-de-BN}) and (\ref{BN-matrix}) we obtain\footnote{Since in the $\bt $-functions we are ignoring
$Y_{\nu}$ couplings (due to their smallness), for all practical purposes the $m_B$ can be treated as a constant.}
\beq
{\rm at}~\mu=M:~~~~B_N=m_BM\left(
  \begin{array}{cc}
    -\al \de_{N}(1+\bar{\ep }_1) & 1 \\
    1 & -\al \de_{N}^*(1+\bar{\ep }_2) \\
  \end{array}
\right),~~~~\al =1+2\fr{m_A}{m_B}~
\la{BN-at-M}
\eeq
and
$$
\bar{\ep }_1\!=\!\fr{1}{4\pi^2\al \de_{N}}\!\int_{t_M}^{t_G} \!\!dt\!\l \!\!Y_{\nu }^\dag (\fr{\al }{16\pi^2}Y_eY_e^\dag +2\fr{m_A}{m_B}\hat{\ep })Y_{\nu }\!\r_{\!\!21}\!,~~~
\bar{\ep }_2\!=\!\fr{1}{4\pi^2\al \de_{N}^*}\!\int_{t_M}^{t_G} \!\!dt\!\l \!\!Y_{\nu }^\dag (\fr{\al^* }{16\pi^2}Y_eY_e^\dag +2\fr{m_A^*}{m_B^*}\hat{\ep }^*)Y_{\nu }\!\r_{\!\!21}^{\!\!*}~,~
$$
\beq
{\rm with}~~~~\hat{\ep }={\rm Diag}\l \ep_0,~ \ep_0,~ \ep_0+\ep \r .
\la{eps-12}
\eeq
The form of $B_N$ given in Eq. (\ref{BN-at-M}) will be needed to construct the sneutrino mass matrix, which we will do below.

\subsection{Sneutrino Mass Matrix and its Diagonalization}

For calculating scalar RHN masses, from (\ref{V-SB-nu}) we keep only $B_N$-term. Also include the mass$^2$ term $\tl N^\dag M_N^\dag M_N\tl N$ coming from the superpotential.
Therefore, we consider the following quadratic potential:
\beq
V_{\tl N}^{(2)}=\tl N^\dag M_N^\dag M_N\tl N-\l \fr{1}{2}\tl N^TB_N\tl N+{\rm h.c. }\r ~.
\la{V-tlN-1}
\eeq
With the transformation of the $N$ superfields $N=U_NN'$ (according to Eq. (\ref{MN-diag-tion}), the $U_N$ diagonalizes the fermionic RHN mass matrix), we obtain:
\beq
V_{\tl N}^{(2)}=\tl N^{\hs{0.5mm}'\!\dag } (M_N^{Diag})^2\tl N'-\l \fr{1}{2}\tl N^{\hs{0.5mm}'T }U_N^TB_NU_N\tl N'+{\rm h.c. }\r ~.
\la{V-tlN-2}
\eeq
On the other hand,  from (\ref{BN-at-M}) we have
\beq
U_N^TB_NU_N=m_B|M|\left(
  \begin{array}{cc}
    1-\tl{\al }|\de_{N}| & \fr{i}{2}\al |\de_{N}|(\bar{\ep }_1-\bar{\ep }_2) \\
   \fr{i}{2}\al |\de_{N}|(\bar{\ep }_1-\bar{\ep }_2) & 1+\tl{\al }|\de_{N}|  \\
  \end{array}
\right),~~~{\rm with}~~~~\tl{\al}=\al (1+\fr{\bar{\ep }_1+\bar{\ep }_2}{2})~.
\la{BN-1}
\eeq
With further phase redefinition
\beq
\tl N'=\tl P_1\tl N''~,~~~~~\tl P_1={\rm Diag}\l e^{-i\tl{\om}_1/2}, e^{-i\tl{\om}_2 /2}\r~,~~~~~{\rm with}~~~\tl{\om}_{1,2}={\rm Arg}[m_B(1\mp \tl{\al}|\de_N|)]
\la{scN-phase}
\eeq
and by going to the real scalar components
\beq
{\tl{N}_1}''=\fr{1}{\sq{2}}(\tl{N}_1^R+i\tl{N}_1^I)~,~~~~~{\tl{N}_2}''=\fr{1}{\sq{2}}(\tl{N}_2^R+i\tl{N}_2^I)~,
\la{tlN-real}
\eeq
we will have
$$
-\l \fr{1}{2}\tl N^{\hs{0.5mm}'T }U_N^TB_NU_N\tl N'+{\rm h.c. }\r =-\fr{|Mm_B|}{2} \left |1-\tl{\al}|\de_N|\right |\l \!(\tl{N}_1^R)^2-(\tl{N}_1^I)^2\!\r
$$
$$
-\fr{|Mm_B|}{2} \left |1+\tl{\al}|\de_N|\right |\l \!(\tl{N}_2^R)^2-(\tl{N}_2^I)^2\!\r
-|M|{\rm Re}(m_B\de_{\ep})\l \tl{N}_1^R\tl{N}_2^R-\tl{N}_1^I\tl{N}_2^I\r
+|M|{\rm Im}(m_B\de_{\ep})\l \tl{N}_1^I\tl{N}_2^R+\tl{N}_1^R\tl{N}_2^I\r
$$
\beq
{\rm with}~~~~\de_{\ep}=i\al |\de_N|\fr{\bar{\ep }_1-\bar{\ep }_2}{2}e^{-i(\tl{\om}_1+\tl{\om}_2)/2}~.
\la{BN-2}
\eeq
From (\ref{V-tlN-2}) and (\ref{BN-2}) we obtain the mass$^2$ terms:
\beq
V_{\tl N}^{(2)}=\fr{1}{2}\tl{n}^{0T}M_{\tl n}^2\tl n^0~,~~~~{\rm with}~~~~\tl{n}^{0T}=\l \tl{N}_1^R, \tl{N}_1^I, \tl{N}_2^R, \tl{N}_2^I\r ~
\la{V-tlN-3}
\eeq
and
\beq
M_{\tl n}^2=
\left(
  \begin{array}{cccc}
    (\tl M_1^0)^2 & 0 & -|M|{\rm Re}(m_B\de_{\ep}) & |M|{\rm Im}(m_B\de_{\ep}) \\
    0 & (\tl M_2^0)^2 & |M|{\rm Im}(m_B\de_{\ep}) & |M|{\rm Re}(m_B\de_{\ep}) \\
    -|M|{\rm Re}(m_B\de_{\ep}) & |M|{\rm Im}(m_B\de_{\ep}) & (\tl M_3^0)^2 & 0 \\
    |M|{\rm Im}(m_B\de_{\ep}) & |M|{\rm Re}(m_B\de_{\ep}) & 0 & (\tl M_4^0)^2 \\
  \end{array}
\right)
\la{Mtln}
\eeq
where
$$
(\tl M_1^0)^2=|M|^2(1-|\de_N|)^2-|m_BM|\left |1-\tl{\al}|\de_N|\right |,~~(\tl M_2^0)^2=|M|^2(1-|\de_N|)^2+|m_BM|\left |1-\tl{\al}|\de_N|\right |~,
$$
\beq
(\tl M_3^0)^2=|M|^2(1+|\de_N|)^2-|m_BM|\left |1+\tl{\al}|\de_N|\right |,~~(\tl M_4^0)^2=|M|^2(1+|\de_N|)^2+|m_BM|\left |1+\tl{\al}|\de_N|\right |
\la{M0s}
\eeq

The coupling of $\tl{n}^0$ states with the fermions emerges from the $F$-term of the superpotential $l^TY_{\nu }Nh_u$.
Following the transformations, indicated above, we will have:
$$
(l^TY_{\nu }Nh_u)_F\to \tl h_ul^TY_{\nu }\tl N=e^{-i\tl{\om}_2/2}\tl h_ul^T Y_{\nu }U_N\l \rho_ue^{i(\tl{\om}_2-\tl{\om}_1)/2},\rho_d\r \tl{n}^0~,
$$
\beq
{\rm with}~~~
\rho_u=\fr{1}{\sq{2}}\l  \begin{array}{cc} 1&i\\0&0 \end{array}\r ~,~~~
\rho_d=\fr{1}{\sq{2}}\l  \begin{array}{cc} 0&0\\1&i \end{array}\r ~.
\la{YF0}
\eeq
Performing the diagonalization of the matrix (\ref{Mtln}) by the transformation $V_{\tl n}^TM_{\tl n}^2V_{\tl n}=(M_{\tl n}^{Diag})^2$,
$\tl{n}^0=V_{\tl n}\tl n$,
the fermion coupling with the scalar  $\tl n$ eigenstates will be
\beq
\tl h_ul^TY_F\tl n~~~~~~{\rm with}~~~Y_F=Y_{\nu }\tl V^0 V_{\tl n}~,
~~~\tl V^0=U_N\l \rho_ue^{-i\tl{\om}_1/2},\rho_de^{-i\tl{\om}_2/2}\r ~.
\la{YF}
\eeq
The coupling with the slepton $\tl l$ is derived from the interaction term $h_u\tl l^T\l Y_{\nu }M_N^*\tl N^*-A_{\nu }\tl N\r $.
Going from  $\tl N$ to the $\tl n$ states, we obtain
\beq
h_u\tl l^TY_B\tl n~~~~{\rm with}~~~Y_B=(Y_{\nu }M_N^*\tl V^{0*}-A_{\nu }\tl V^0)V_{\tl n}~.
\la{YB}
\eeq
For a given values of $M, m_B$ and $m_A$, with help of Eqs. (\ref{Mtln}), (\ref{YF}) and (\ref{YB}), we will have coupling matrices $Y_F$, $Y_B$ and all other quantities needed for calculation of the baryon asymmetry
created via the decays of the $\tl n_{1,2,3,4}$ states.

\subsection{Asymmetry Via $\tl{n}$ Decays }

Now we are ready to discuss the contribution to the net baryon asymmetry from the out of equilibrium resonant decays of the right
handed sneutrinos (RHS).
As we have seen, with SUSY breaking terms, the masses of RHS's  differ from their fermionic partners' masses. Thus we have mass-eigenstate RHS's $\tl{n}_{i=1,2,3,4}$
with masses $\tl{M}_{i=1,2,3,4}$ respectively.  With
the SUSY scale $M_S$ smaller (at least by a factor of 3) than the scale $M$, the states $\tl n_i$ remain nearly degenerate.

 For
the resonant $\tl{n}$-decays we will apply resummed effective amplitude technique \cite{Pilaftsis:1997jf}. Effective amplitudes
for the real $\tl{n}_i$ decay, say into the lepton $l_{\al }$ ($\al=1,2,3$) and antilepton $\ov{l}_{\al }$ respectively are given by \cite{Pilaftsis:1997jf}
\beq
\hat{S}_{\al i}=S_{\al i}-\sum_jS_{\al j}\fr{\Pi_{ji}(\tl{M}_i)(1-\de_{ij})}{\tl{M}_i^2-\tl{M}_j^2+\Pi_{jj}(\tl{M}_i)}~,~~
\hat{\ov{S}}_{\al i}=S_{\al i}^*-\sum_jS_{\al j}^*\fr{\Pi_{ji}(\tl{M}_i)(1-\de_{ij})}{\tl{M}_i^2-\tl{M}_j^2+\Pi_{jj}(\tl{M}_i)}~,
\la{eff-amps}
\eeq
where $S_{\al i}$ is a tree level amplitude and $\Pi_{ij}$ is a two point Green function's (polarization operator of $\tl{n}_i-\tl{n}_j$)
absorptive part. The CP asymmetry is then given by
\beq
\ep_i^{sc}=\fr{\sum_{\al }\l |\hat{S}_{\al i}|^2-|\hat{\ov{S}}_{\al i}|^2\r}{\sum_{\al }\l |\hat{S}_{\al i}|^2+|\hat{\ov{S}}_{\al i}|^2\r }~.
\la{asym-byS}
\eeq

With $Y_F$ and $Y_B$ given by Eq. (\ref{YF}) and (\ref{YB}) we can calculate  polarization diagram's (with external legs $\tl{n}_i$ and $\tl{n}_j$) absorptive part $\Pi_{ij}$, which at the 1-loop level  is given by:
\beq
\Pi_{ij}(p)=\fr{i}{8\pi}\l p^2Y_F^\dag Y_F+p^2Y_F^T Y_F^*+Y_B^\dag Y_B+Y_B^TY_B^*\r_{ij} ~,
\la{abs-part}
\eeq
where $p$ denotes external momentum  in the diagram and upon evaluation  of (\ref{asym-byS}), for
 $\Pi $ we should use (\ref{abs-part}) with $p=\tl{M}_i$.

%
%
%
\begin{table}
\vs{0.1cm}
\hs{2.68cm}
\begin{tabular}{|c|c|c|}
  \hline
  ~~~~~~~  & \!$(m_A, m_B)=(100~\!i, 500) {\rm GeV}$\! & \!$(m_A, m_B)\!=\!(500, 1000) {\rm GeV}$ \! \\
  \hline
\end{tabular}
\vs{-0.45cm}
 $$\begin{array}{|c|c|c|c|c|}
\hline
\vs{-0.3cm}
&&&&\\
{\rm Case}  & ~~~10^{4}\!\tm \!a_2~~~~& ~~~10^{11}\!\tm \!\fr{\tl n_b}{s}~~~ & ~~~ 10^{4}\!\tm \!a_2~~~~ &~~~10^{11}\!\tm \!\fr{\tl n_b}{s}~~~\\
\hline
{\bf (I_{-})} & 0.016 & 0.25 & 0.016& 0.24\\
\hline
 {\bf (I.1)} & 0.0159785 & 0.25 &0.0159785 &0.25  \\
 \hline
 {\bf (I.2)} &  0.0299 &0.24 &0.0299 & 0.24\\
\hline
 {\bf (I.3)} & 0.0987  &0.24 & 0.0987& 0.24\\
 \hline
 {\bf (I.4)} & 0.3237 &0.24 &0.3237 &0.24\\
 \hline
 {\bf (I.5)} &1.05655 & 0.23 &1.05655 &0.23\\
 \hline
 \hline
 {\bf (II_{-})} &0.0229 &0.25 &0.0229 & 0.24\\
 \hline
 {\bf (II.1)} &0.0229 &0.25 & 0.0229& 0.24\\
 \hline
 {\bf (II.2)} & 0.02986 & 0.24& 0.02986& 0.24\\
 \hline
 {\bf (II.3)} &0.09835 &0.24  & 0.09835& 0.24 \\
 \hline
 {\bf (II.4)} & 0.322 & 0.24&0.322 & 0.24\\
 \hline
 {\bf (II.5)} & 1.05 & 0.23& 1.05& 0.23\\
 \hline
\end{array}$$
\vs{-0.5cm}
\caption{Values of $\fr{\De n_b}{s}=\fr{\tl n_b}{s}$ - contributions to the Baryon asymmetry via decays of the right handed sneutrinos
 for cases given in table \ref{results-for-P1} (i.e. for values of $a_2$ giving $\l \fr{n_b}{s}\!\r_{\rm max}$ given in Tab. \ref{results-for-P1}). These values correspond to the phases $\de =-0.378$ and $\phi_{+}=-1.287$.  }
 \vs{-0.3cm}
 \label{scal-asym-for-P1}
\end{table}

In an unbroken SUSY limit, neglecting finite temperature effects ($T\to 0$),
the $\tl{N}$ decay does not produce lepton asymmetry due to the  following reason. The decays of $\tl{N}$ in the fermion and scalar channels are respectivelly
 $\tl{N}\to l\tl{h}_u$ and $\tl{N}\to \tl{l}^*h_u^*$. Since the rates of these processes are the same due to SUSY  (at $T=0$),
 the lepton asymmetries created from these decays cancel each other.
 With  $T\neq 0$, the cancelation does not take place  and one has
\beq
\tl{\ep }_i=\ep_i(\tl{n}_i\to l\tl{h}_u)\De_{BF}~,
\la{ep-non-zero}
\eeq
with a temperature dependent factor $\De_{BF}$ given in \cite{DAmbrosio:2003nfv}.\footnote{
This expression is valid with  alignment $A_{\nu }=m_AY_{\nu }$, which we are assuming at the GUT scale and thus Eq. (\ref{ep-non-zero})
can be well applicable for our estimates.}
Therefore, we just need to compute $\ep_i(\tl{n}_i\to l\tl{h}_u)$, which is the asymmetry created by $\tl{n}_i$ decays in two fermions.
Thus, in (\ref{eff-amps}) we take $S_{\al i}=(Y_F)_{\al i}$ and calculate $\ep_i(\tl{n}_i\to l\tl{h}_u)$ with (\ref{asym-byS}).
The baryon asymmetry created from the lepton asymmetry due to $\tl{n}$ decays is:
\beq
\fr{\tl{n}_b}{s}\simeq -8.46\cdot 10^{-4}\sum_{i=1}^4\fr{\tl{\ep }_i}{\De_{BF}}\eta_i=
-8.46\cdot 10^{-4}\sum_{i=1}^4\ep_i(\tl{n}_i\to l\tl{h}_u)\eta_i~,
\la{sc-asym}
\eeq
where an effective number of degrees of freedom (including two RHN superfields) $g_*=228.75$ was used.
$\eta_i$ are efficiency factors which depend on $\tl{m}_i\simeq \fr{(v\sin \bt )^2}{M}2(Y_F^\dag Y_F)_{ii}$, and take into account temperature effects by integrating the Boltzmann equations \cite{DAmbrosio:2003nfv}.

In table \ref{scal-asym-for-P1} we give results for the neutrino model discussed in Sect. \ref{sect-P1-texture}. These are obtained for the SUSY particle masses $=M_S$ and for the different values of pairs $(m_A, m_B)$ (see also the caption of the table \ref{scal-asym-for-P1}).
Upon the calculations,
with obtained values
of $\tl{m}_i$, according to Ref. \cite{DAmbrosio:2003nfv} we picked up the corresponding values of $\eta_i$ and
used them in  (\ref{sc-asym}).
From table \ref{scal-asym-for-P1} we see that contribution to the net baryon asymmetry from the RHS decays is
suppressed $\fr{\tl{n}_b}{n_b}<3\cdot 10^{-2}$, i.e. is less than $3\%$.


\bibliographystyle{unsrt}

\end{document}